\documentclass[twocolumn,secnumarabic,amssymb,amsmath, nobibnotes, aps, prd, superscriptaddress]{revtex4-2}

\usepackage{physics}
\usepackage{mathtools} 
\usepackage{comment} 
\usepackage{graphicx}
\usepackage[hidelinks]{hyperref}

\usepackage{dcolumn}
\usepackage{threeparttable}
\newtheorem{thm}{Thm.}

\newtheorem{lem}[thm]{Lem.}

\begin{document}
\title{Quantum master equation for many-body systems based on the Lieb-Robinson bound}%

\author{Koki Shiraishi}%
\email{shiraishi@cat.phys.s.u-tokyo.ac.jp}
\affiliation{
  Department of Physics, The University of Tokyo, 7-3-1 Hongo, Bunkyo-ku, Tokyo 113-0033, Japan.
}
\author{Masaya Nakagawa}%
\affiliation{
  Department of Physics, The University of Tokyo, 7-3-1 Hongo, Bunkyo-ku, Tokyo 113-0033, Japan.
}
\author{Takashi Mori}%
\affiliation{
  Department of Physics, Keio University, Kohoku-ku, Yokohama, Kanagawa 223-8522, Japan.
}
\author{Masahito Ueda}%
\affiliation{
  Department of Physics, The University of Tokyo, 7-3-1 Hongo, Bunkyo-ku, Tokyo 113-0033, Japan.
}
\affiliation{RIKEN Center for Emergent Matter Science (CEMS), Wako, Saitama 351-0198, Japan.}
\affiliation{Institute for Physics of Intelligence, The University of Tokyo, 7-3-1 Hongo, Bunkyo-ku, Tokyo 113-0033, Japan.}
\date{\today}%
\begin{abstract}
  The local Gorini-Kossakowski-Sudarshan-Lindblad (GKSL) quantum master equation is a powerful tool for the study of open quantum many-body systems.
  However, its microscopic derivation applicable to many-body systems is available only in limited cases, and it has yet to be fully understood under what microscopic conditions the local GKSL equation is valid.
  We derive the local GKSL equation on the basis of the Lieb-Robinson bound, which provides an upper bound of the propagation of information in a broad class of quantum many-body systems.
  We numerically test the validity of the derived local GKSL equation for a one-dimensional tight-binding fermion chain.
\end{abstract}
\maketitle

\section{Introduction}
\label{sec:introduction}
Most quantum systems are inevitably influenced by surrounding environments, i.e., they are open quantum systems~\cite{breuer2002theory}.
In recent years, open quantum many-body systems have intensively been studied both theoretically and experimentally due to advances in experimental techniques~\cite{AJDaley,MULLER20121}.
The standard method to describe an open quantum system is based on the quantum master equations (QMEs)~\cite{10.1143/PTP.20.948,doi:10.1063/1.1731409,PhysRev.89.728,5392713,REDFIELD19651,doi:10.1063/1.522979,Lindblad1976}, such as the Gorini-Kossakowski-Sudarshan-Lindblad (GKSL) equation~\cite{doi:10.1063/1.522979,Lindblad1976} and the Redfield equation~\cite{REDFIELD19651}.
In fact, the QMEs can correctly reproduce many experimental results about open quantum many-body systems~\cite{barreiro2011open,PhysRevLett.110.035302,PhysRevX.7.011034,doi:10.1126/sciadv.1701513,bouganne2020anomalous,PhysRevX.11.041046}.
The QMEs have also been used to analyze transport phenomena in nonequilibrium steady states of quantum systems in contact with reservoirs~\cite{PhysRevB.80.035110,PhysRevLett.106.217206,RevModPhys.94.045006,michel2003fourier,mejia2007heat,Steinigeweg_2009,PhysRevE.86.061118,Manzano_2016}.

Among the QMEs, the local GKSL equation is widely applied to open many-body system~\cite{PhysRevE.76.031115,PhysRevB.105.115139,PhysRevLett.106.217206,PhysRevLett.107.137201,RevModPhys.93.025003,AJDaley,RevModPhys.94.045006}.
The GKSL equation is given by (with $\hbar=1$)
\begin{equation}
  \label{eq:GKSL}
  \frac{d}{dt}\rho=-i[H,\rho]+\sum_\mu \left(L_\mu \rho L_\mu^\dagger-\frac{1}{2}\{L_\mu^\dagger L_\mu,\rho\}\right),
\end{equation}
where $\rho$ is the density matrix of a system, $H$ is the Hamiltonian of the system, and $L_\mu$'s are the Lindblad operators.
Equation~(\ref{eq:GKSL}) is called the local GKSL equation if each Lindblad operator in Eq.~(\ref{eq:GKSL}) acts only on a spatially local subsystem.
The local GKSL equation reflects the local nature of the effect of dissipation due to the environment and consequently guarantees the physically desirable properties.
For example, the locality of the Lindblad operators guarantees the existence of the Lieb-Robinson bound in open quantum many-body systems~\cite{PhysRevLett.104.190401},
thermalization of bulk-dissipated systems in the weak coupling regime~\cite{PhysRevE.101.042116},
and local conservation laws in the bulk of boundary-dissipated many-body systems~\cite{PhysRevA.105.032208}.

However, since the local GKSL equation is usually given phenomenologically without any microscopic derivation~\cite{RevModPhys.94.045006,PhysRevA.107.062216}, it is not clear how properties such as the temperature of the environment should be incorporated in the Lindblad operators.
It is also unclear when the local GKSL equation is valid.
The validity of the local GKSL equation is debated in terms of thermodynamics~\cite{Levy_2014,Barra2015,Hofer_2017,PhysRevE.97.022115,DeChiara_2018,RevModPhys.93.035008,Scali2021localmaster}, conservation laws~\cite{PhysRevA.105.032208}, dynamics~\cite{Scali2021localmaster} and phase transitions~\cite{PhysRevResearch.4.013171}.
The microscopic derivations of the local GKSL equation known so far are applicable to limited cases,
such as a case of the collisional bath~\cite{PhysRev.129.1880,Englert2002,PhysRevA.65.040102,PhysRevLett.88.097905,PhysRevLett.102.207207,Ciccarello+2017+53+63,CICCARELLO20221} and a case in which the intersite couplings are assumed to be weak enough to be treated perturbatively~\cite{PhysRevE.76.031115,Trushechkin_2016,schnell2023global}.
Therefore, a microscopic derivation of the local GKSL equation applicable to generic many-body systems is highly desired.

In this paper, we propose a method to derive the local GKSL equation by using the Lieb-Robinson bound~\cite{Lieb1972,doi:10.1137/18M1231511}, which provides a fundamental limitation on the speed of information propagation in locally interacting quantum many-body systems.
The GKSL equation is commonly derived from the Redfield equation by using the Born-Markov approximation and further approximations in the weak-coupling regime, where the system-environment interaction is weak~\cite{breuer2002theory}.
Here, we use the Lieb-Robinson bound to impose locality on the Redfield equation before applying existing approximations to obtain the GKSL equation such as the rotating-wave (secular) approximation~\cite{Davies1974,doi:10.1063/1.4907370,https://doi.org/10.48550/arxiv.1710.09939,PhysRevA.100.012107}, time coarse-graining~\cite{Cohen-Tannoudji_1986,LIDAR200135,PhysRevA.78.022106,PhysRevA.79.032110,PhysRevA.88.012103,Mozgunov2020completelypositive}, and approximation of the sum of the spectral densities by the product of the square roots of the spectral densities~\cite{PhysRevB.97.035432,PhysRevB.102.115109,Davidovic2020completelypositive}.
On the basis of this microscopic derivation, we find that the Lindblad operators of the local GKSL equation should have a support of size $\zeta_0 \tau_B$, where $\zeta_0$ is the propagation velocity of the system and $\tau_B$ is the relaxation time of the environment (their precise definitions are given later).
We also numerically demonstrate that the local GKSL equation can correctly describe the steady states of open many-body systems when the relaxation of the environment is sufficiently fast.

Our microscopic derivation clarifies how the locality of the Lindblad operators is related to the time scales of the system, the environment, and the interaction between them.
Moreover, our derivation reduces the computational cost of time evolution since the Lindblad operators can be calculated only from diagonalization of the Hamiltonian of a local subsystem.
The use of local GKSL equations based on the microscopic derivation is expected to deepen understanding of nonequilibrium phenomena in open quantum many-body systems.

The rest of this paper is organized as follows.
In Sec.~\ref{sec:2}, we introduce the Redfield equation and two types of the GKSL equations, namely, the Davies equation~\cite{Davies1974} and the Nathan-Rudner equation~\cite{PhysRevB.102.115109}.
In Sec.~\ref{sec:3}, we derive the local GKSL equation on the basis of the Lieb-Robinson bound.
In Sec.~\ref{sec:4}, we numerically evaluate the error of our estimates about the equilibrium and nonequilibrium steady states described by the derived local GKSL equation for a quadratic fermionic system and show that the error becomes small when the parameter regime is consistent with our derivation based on the Lieb-Robinson bound.
In Sec.~\ref{sec:5}, we discuss the numerical cost for the analysis of open quantum many-body systems on the basis of the derived local GKSL equation.
Finally, we conclude this paper in Sec.~\ref{sec:6}.

In Appendix~\ref{sec:appendixA}, we give the derivation of the Nathan-Rudner equation in the frequency domain.
In Appendix~\ref{sec:appendixBB}, we show that the largest eigenvalue of the dissipator of the QME in the quadratic open fermionic system can be efficiently calculated.
In Appendix~\ref{sec:appendixB}, we provide additional numerical results.

\section{Quantum master equations}
\label{sec:2}
In this section, we introduce three QMEs, namely, the Redfield equation, the Davies equation, and the Nathan-Rudner equation (NRE).
We summarize in Table~\ref{tab:table1} the conditions under which these and two related QMEs are valid, including the local GKSL equations in the next section.

\begin{table}[b]
  \begin{ruledtabular}
  \begin{tabular}{lc}
  \textrm{QME}&
  \textrm{Conditions}\\
  \colrule
  \textrm{Redfield}~\cite{REDFIELD19651,breuer2002theory} & $\tau_\mathrm{B}\ll \tau_\mathrm{SB}$ \\
  \textrm{Davies}~\cite{Davies1974,breuer2002theory} & $\tau_\mathrm{B}\ll \tau_\mathrm{SB},\tau_\mathrm{S}\ll \tau_\mathrm{SB}$ \\
  \textrm{NRE}~\cite{PhysRevB.102.115109,Davidovic2020completelypositive} & $\tau_\mathrm{B}\ll \tau_\mathrm{SB}$ \\
  \textrm{Local Davies} & $\tau_\mathrm{B}\ll \tau_\mathrm{SB}$, $\tau_\mathrm{B}\ll R/\zeta_0$\footnotemark[1] , $\tau_\mathrm{S,\Omega}\ll \tau_\mathrm{SB}$\footnotemark[2] \\
  \textrm{Local NRE} & $\tau_\mathrm{B}\ll \tau_\mathrm{SB}$, $\tau_\mathrm{B}\ll R/\zeta_0$\footnotemark[1] \\
  \end{tabular}
  \end{ruledtabular}
  \footnotetext[1]{Discussed in Sec.~\ref{sec:LRF}. Numerically shown in Sec.~\ref{sec:4}.}
  \footnotetext[2]{Discussed in Sec.~\ref{sec:locDavies}. Numerically shown in Sec.~\ref{sec:tradeoff}.}
  \caption{\label{tab:table1}
  Conditions required for  various QMEs to be valid, where
  $\tau_\mathrm{B}$, $\tau_\mathrm{SB}$, $\tau_\mathrm{S}$, and $\tau_{\mathrm{S},\Omega}$ are
  the relaxation time of the bath,
  the time scale of the system-bath interaction,
  the time scale of the system,
  and the time scale of the local subsystem $\Omega$, respectively,
  $R$ is the radius of $\Omega$,
  and $\zeta_0$ is the propagation velocity of the system.
  The detailed definitions are given in the text.
  }
  \end{table}

Let us consider a quantum system $\mathrm{S}$, which interacts with a bath $\mathrm{B}$ and is represented by a finite-dimensional Hilbert space.
Since we consider a many-body system, the dimension of the Hilbert space grows exponentially with increasing the system size.
The Hamiltonian of the total system is given by $H_\mathrm{tot}=H_\mathrm{S}\otimes I_\mathrm{B}+I_\mathrm{S}\otimes H_\mathrm{B}+H_\mathrm{SB}$,
where $H_\mathrm{S}$, $H_\mathrm{B}$, and $H_\mathrm{SB}$ are the Hamiltonians of the system, the bath, and the system-bath interaction, respectively, and $I_{\mathrm{S}(\mathrm{B})}$ is the identity operator on the system (bath).
The interaction Hamiltonian is represented by
\begin{equation}
  \label{eq:SBint}
  H_\mathrm{SB}=\sum_\mu A_\mu\otimes B_\mu,
\end{equation}
where $A_\mu$'s and $B_\mu$'s are operators that act on the system and the bath, respectively.
The time evolution of the total system is described by the von-Neumann equation
\begin{equation}
  \label{eq:vN}
  \frac{d}{dt}\rho_\mathrm{tot}(t)=-i[H_\mathrm{tot},\rho_\mathrm{tot}(t)],
\end{equation}
where $\rho_{\mathrm{tot}}(t)$ is the density matrix of the total system at time $t$.

\subsection{Redfield equation}
We first introduce the Redfield equation, from which the GKSL equation can be derived.
We assume that the interaction between the system and the bath is weak and that the state $\rho_\mathrm{B}$ of the bath is the Gibbs state $\rho_\mathrm{B}=e^{-\beta H_\mathrm{B}} / \mathrm{Tr}[e^{-\beta H_\mathrm{B}}]$ with inverse temperature $\beta$ throughout time evolution.
The correlation functions of operators $B_\mu$ of the bath are denoted by $C_{\mu\nu}(t)=\tr(B_\mu^{\dagger}(t)B_\nu \rho_\mathrm{B})$, where $B_\mu(t)=e^{iH_\mathrm{B}t}B_\mu e^{-iH_\mathrm{B}t}$.

We introduce the time scale $\tau_\mathrm{B}$ of the bath and the time scale $\tau_\mathrm{SB}$ of the system-bath interaction.
The precise definitions of the time scales are given below.
If the system-bath interaction is so weak and the relaxation of the bath is so fast that there is a separation of the two time scales such that $\tau_\mathrm{B}\ll\tau_\mathrm{SB}$, 
then the Born-Markov approximation~\cite{breuer2002theory,REDFIELD19651} is justified and the Redfield equation
\begin{equation}
  \label{eq:RedfieldSchr}
  \begin{split}
    \frac{d}{dt}\rho(t)=&-i[H_\mathrm{S},\rho(t)]\\
    &+\sum_{\mu,\nu}\int_0^\infty{ds}\left[ C_{\mu\nu}(s)(A_\nu(-s)\rho(t)A_\mu^{\dagger}\right.\\
    &~~~~~~~~~~~~~~~~\left.-A_\mu^{\dagger} A_\nu(-s)\rho(t))+\mathrm{H.c.}\right]
  \end{split}
\end{equation}
can be derived from Eq.~(\ref{eq:vN})~\cite{REDFIELD19651}.
Here, $\rho(t)=\mathrm{tr}_{\mathrm{B}}[\rho_{\mathrm{tot}}(t)]$ is the reduced density matrix of the system and we assume the condition $\tr_\mathrm{B}[H_\mathrm{SB},\rho(0)\otimes\rho_\mathrm{B}]=0$ for the initial state $\rho(0)$ of the system.
Let $E_n$ and $\ket{E_n}$ be an eigenvalue of $H_{\mathrm{S}}$ and the corresponding eigenstate.
By decomposing $A_\mu(t)$ into the sum of $A_\mu(\omega)$'s given by
\begin{equation}
  \label{eq:FreqDecomp}
  A_\mu(\omega)=\sum_{E_n-E_m=\omega}\ket{E_m}\bra{E_m}A_\mu\ket{E_n}\bra{E_n},
\end{equation}
we can rewrite the Redfield equation~(\ref{eq:RedfieldSchr}) as
  \begin{equation}
    \label{eq:frequencyMRSchr}
    \begin{split}
      \frac{d}{dt}&\rho(t)=-i[H_\mathrm{S}+H_\mathrm{LS},\rho(t)]\\
      &+\sum_{\mu,\nu}\sum_{\omega,\omega^\prime}\left[\frac{\gamma_{\mu\nu}(\omega)+\gamma_{\mu\nu}(\omega^\prime)}{2}+i(\eta_{\mu\nu}(\omega)-\eta_{\mu\nu}(\omega^\prime))\right]\\
      &\times\left(A_\nu(\omega)\rho(t){A_\mu}^\dagger(\omega^\prime)-\frac{1}{2}\{{A_\mu}^\dagger(\omega^\prime){A_\nu}(\omega),\rho(t)\}\right),
    \end{split}
  \end{equation}
where the Lamb-shift Hamiltonian $H_\mathrm{LS}$ is defined as
\begin{equation}
  \begin{split}
    H_\mathrm{LS}=&\sum_{\mu,\nu}\sum_{\omega,\omega^\prime}\left(\frac{\eta_{\mu\nu}(\omega)+\eta_{\mu\nu}(\omega^\prime)}{2}\right.\\
    &\left.+i\frac{\gamma_{\mu\nu}(\omega)-\gamma_{\mu\nu}(\omega^\prime)}{4}\right){A_\mu}^\dagger(\omega^\prime){A_\nu}(\omega).
  \end{split}
\end{equation}
Here, the power spectrum $\gamma_{\mu\nu}(\omega)$ and the principal value integral $\eta_{\mu\nu}(\omega)$ of the power spectrum satisfy
\begin{equation}
  \Gamma_{\mu\nu}(\omega)\coloneqq\int_{0}^\infty{ds}e^{i\omega s}C_{\mu\nu}(s)=\frac{1}{2}\gamma_{\mu\nu}(\omega)+i\eta_{\mu\nu}(\omega).
\end{equation}
The power spectrum  $\gamma_{\mu\nu}(\omega)$ is given by
\begin{equation}
  \gamma_{\mu\nu}(\omega)=\int_{-\infty}^\infty{ds}e^{i\omega s}C_{\mu\nu}(s),
\end{equation}
and the matrices $\gamma(\omega)$ and $\eta(\omega)$ are Hermitian: $\gamma_{\nu\mu}^*(\omega)=\gamma_{\mu\nu}(\omega), \eta_{\nu\mu}^*(\omega)=\eta_{\mu\nu}(\omega)$.

The time scale $\tau_\mathrm{B}$ of the bath is defined by the relaxation time of the correlation functions as
\begin{equation}
  \label{eq:tB}
  \tau_\mathrm{B}=\max_{\mu,\nu}\frac{\int_0^\infty t|C_{\mu\nu}(t)|dt}{\int_0^\infty |C_{\mu\nu}(t)|dt},
\end{equation}
where $|C_{\mu\nu}|$ is the absolute value of $C_{\mu\nu}$~\cite{Mozgunov2020completelypositive,PhysRevB.102.115109}.
The time scale $\tau_\mathrm{SB}$ of the system-bath interaction is defined as the inverse rate of the time evolution due to the system-bath interaction.
Let $\mathcal{D}[\rho(t)]$ denote the second term in Eq.~(\ref{eq:RedfieldSchr}) as
\begin{equation}
  \label{eq:dissipator}
  \begin{split}
    \mathcal{D}[\rho(t)]=&\sum_{\mu,\nu}\int_0^\infty{ds}\left[ C_{\mu\nu}(s)(A_\nu(-s)\rho(t)A_\mu^{\dagger}\right.\\
      &~~~~~~~~~~~~~~~~\left.-A_\mu^{\dagger} A_\nu(-s)\rho(t))+\mathrm{H.c.}\right],
  \end{split}
\end{equation}
which represents the time evolution caused by the system-bath interaction.
The rate of the evolution caused by $\mathcal{D}[\rho(t)]$ is bounded from above by the operator norm of $\mathcal{D}$:
\begin{equation}
  \label{eq:tSB}
  \|\mathcal{D}[\rho(t)]\|_1\leq \|\mathcal{D}\|\|\rho(t)\|_1\leq  \|\mathcal{D}\|=:\tau_\mathrm{SB}^{-1},
\end{equation}
where $\|\cdot\|$ is the operator norm induced by the trace norm $\|\cdot\|_1$.
We note that ${\tau_\mathrm{SB}}^{-1}$ only gives an upper bound on the rate of time evolution caused by the system-bath interaction, but does not necessarily characterize the rate of time evolution itself.
Therefore, $\tau_\mathrm{B}\ll\tau_\mathrm{SB}$ is a sufficient condition for deriving the Redfield equation, but not a necessary condition.

Equation~(\ref{eq:frequencyMRSchr}) is not in the GKSL form and does not possess the complete positivity because the Hermitian matrix $\gamma_{\nu\omega,\mu\omega^\prime}\coloneqq (\gamma_{\mu\nu}(\omega)+\gamma_{\mu\nu}(\omega^\prime))/2 + i(\eta_{\mu\nu}(\omega)-\eta_{\mu\nu}(\omega^\prime))$ is not necessarily positive semidefinite.
To recover the complete positivity, we need further approximations to the Redfield equation to derive the GKSL equations.
Such derivations have been studied in Refs.~\cite{Davies1974,Cohen-Tannoudji_1986,LIDAR200135,PhysRevA.78.022106,PhysRevA.79.032110,PhysRevA.88.012103,Mozgunov2020completelypositive,PhysRevB.97.035432,PhysRevB.102.115109,Davidovic2020completelypositive,doi:10.1063/1.4907370,https://doi.org/10.48550/arxiv.1710.09939,PhysRevA.100.012107,PhysRevE.104.014110,PhysRevE.76.031115,PhysRevA.102.032207,Trushechkin_2016}.
In the following, we introduce two types of the microscopically derived GKSL equations that we use in this paper.

\subsection{Davies equation}

The Davies equation~\cite{Davies1974} is the well-known GKSL equation derived by applying the rotating-wave approximation to the Redfield equation.
To derive the Davies equation, we assume that the gaps between energy-level spacings are so large that the time scale $\tau_\mathrm{S}$ of the system defined by a typical value of $|\omega-\omega^\prime|^{-1}$ in Eq.~(\ref{eq:frequencyMRSchr}) is much smaller than $\tau_\mathrm{SB}$.
A typical value is, for example, defined as the maximum value of $|\omega-\omega^\prime|^{-1}$ or the maximum value of $|\omega-\omega^\prime|^{-1}$ over $\omega$ and $\omega^\prime$ satisfying that $\gamma_{\mu\nu}(\omega)$ or $\gamma_{\mu\nu}(\omega^\prime)$ is sufficiently large.
Here, we do not fix the definition of $\tau_{\mathrm{S}}$ and only assume $\tau_{\mathrm{S}} \ll \tau_{\mathrm{SB}}$.
Then, we can neglect the rapidly oscillating terms in Eq.~(\ref{eq:frequencyMRSchr}) where $\omega\neq\omega^\prime$ and obtain the Davies equation
\begin{equation}
  \label{eq:RWALind}
  \begin{split}
    &\frac{d}{dt}\rho (t)=-i[H_\mathrm{S}+H_\mathrm{LS},\rho (t)]
    +\sum_\omega\sum_{\mu,\nu}\gamma_{\mu\nu}(\omega)\\
    &\times\left(A_\nu(\omega)\rho(t) A_\mu^\dagger(\omega)-\frac{1}{2}\{A_\mu^\dagger(\omega)A_\nu(\omega),\rho (t)\}\right),
  \end{split}
\end{equation}
where $H_\mathrm{LS}$ is the Lamb-shift Hamiltonian given by
\begin{equation}
  H_\mathrm{LS}\coloneqq\sum_\omega\sum_{\mu,\nu}\eta_{\mu\nu}(\omega)A_\mu^\dagger(\omega)A_\nu(\omega).
\end{equation}
Since the Lindblad operators of the Davies equation cause quantum jumps between energy eigenstates, the Davies equation (\ref{eq:RWALind}) leads to a global change of the state and therefore does not have desired locality in many-body systems.
Moreover, the rotating-wave approximation often fails for many-body systems~\cite{PhysRevE.76.031115,doi:10.1146/annurev-conmatphys-040721-015537} because the energy levels of many-body systems become exponentially small with increasing the system size (i.e. the condition $\tau_\mathrm{S}\ll\tau_\mathrm{SB}$ is violated).

\subsection{Nathan-Rudner equation}
\label{sec:ULE}
The Nathan-Rudner equation (NRE)~\cite{PhysRevB.102.115109} can be derived under the same assumptions as those made to derive the Redfield equation and it can describe nonequilibrium steady states and dynamics with a small error of $\order{\tau_\mathrm{B}/\tau_\mathrm{SB}}$~\cite{PhysRevB.102.115109,https://doi.org/10.48550/arxiv.2206.02917}.
See Refs.~\cite{PhysRevB.102.115109,Davidovic2020completelypositive} for the detailed derivations.
We also give a derivation of the NRE in Appendix~\ref{sec:appendixA} for the sake of self-containedness of this paper.
Let us define the Lindblad operators $L_\lambda$ and the Lamb-shift Hamiltonian $H_\mathrm{LS}$ as
\begin{equation}
\label{eq:ULELindbladop}
L_\lambda\coloneqq\sum_\mu\sum_\omega \gamma^{1/2}_{\lambda\mu}(\omega)A_\mu(\omega),
\end{equation}
\begin{equation}
  \label{eq:ULEHLS}
  \begin{split}
    H_\mathrm{LS}\coloneqq&\sum_{\mu,\nu}\sum_{\omega,\omega^\prime}\left[\eta_{\mu\nu}\left(\frac{\omega+\omega^\prime}{2}\right)\right.\\
    &+\left.i\frac{\gamma_{\mu\nu}(\omega)-\gamma_{\mu\nu}(\omega^\prime)}{4}\right]A_\mu^\dagger(\omega)A_\nu(\omega^\prime),
  \end{split}
\end{equation}
where $\gamma_{\mu\nu}^{1/2}(\omega)$ is defined to satisfy $\gamma_{\mu\nu}(\omega)=\sum_\lambda\gamma_{\mu\lambda}^{1/2}(\omega)\gamma_{\lambda\nu}^{1/2}(\omega)$.
Then, we can write the NRE in the following form:
\begin{equation}
    \label{eq:ULE}
    \begin{split}
      \frac{d}{dt}\rho=&-i[H_\mathrm{S}+H_\mathrm{LS},\rho ]\\
      &+\sum_{\lambda=1}^k\left(L_\lambda\rho  L_\lambda^\dagger -\frac{1}{2}\{L_\lambda^\dagger L_\lambda,\rho \}\right).
    \end{split}
\end{equation}
The existence of $\gamma_{\mu\nu}^{1/2}(\omega)$ is ensured by the positive semidefiniteness of the matrix $\gamma_{\mu\nu}(\omega)$ at any fixed $\omega$.

The Lindblad operators (\ref{eq:ULELindbladop}) of the NRE  can also be written as
\begin{equation}
\label{eq:ULELindoptime}
L_\lambda=\sum_\mu \int^\infty_{-\infty}ds g_{\lambda\mu}(-s)A_\mu(s),
\end{equation}
where $g_{\lambda\nu}(s)$ is defined as the Fourier transform of $\gamma_{\lambda\nu}^{1/2}(\omega)$:
\begin{equation}
g_{\lambda\mu}(s)\coloneqq\frac{1}{2\pi}\int^\infty_{-\infty}\gamma_{\lambda\mu}^{1/2}(\omega)e^{i\omega s}d\omega.
\end{equation}

\section{Local GKSL equation based on the Lieb-Robinson bound}
\label{sec:3}
In this section, we use the Lieb-Robinson bound to derive the local GKSL equations, which are useful for the analysis of open many-body systems.
We first introduce the Lieb-Robinson bound in Sec.~\ref{sec:LRbound}, and use it to impose the locality on the Redfield equation in Sec.~\ref{sec:LRF}.
We use this local Redfield equation to derive the local GKSL equations, which include the local Davies equation in Sec.~\ref{sec:locDavies} and the local Nathan-Rudner equation in Sec.~\ref{sec:LULE}.

\subsection{Lieb-Robinson bound}
\label{sec:LRbound}
In the following, we consider a many-body system on a lattice with Hamiltonian $H_\mathrm{S}=\sum_{X\subseteq\Lambda}h_X$, where $\Lambda$ denotes the set of sites and $h_X$ is an operator that acts nontrivially only on a local region $X\subseteq\Lambda$.
A distance $\mathrm{dist}(p,q)$ between lattice sites $p$ and $q$ is defined by the number of sites in the shortest pass from site $p$ to site $q$.
We also define the distance between sets $X$ and $Y$ of sites by $\mathrm{dist}(X,Y)\coloneqq\min_{p\in X,q\in Y}\mathrm{dist}(p,q)$.
We assume the Hamiltonian $H_\mathrm{S}$ to be strictly local, that is, $h_X=0$ holds for $X$ whose radius exceeds a certain constant value.

We follow Ref.~\cite{doi:10.1137/18M1231511} to introduce the Lieb-Robinson bound in the form of Lemma~\ref{lem:LRbound}.
\begin{lem}\label{lem:LRbound}
  Let $H=\sum_Yh_Y$ be a local Hamiltonian, and $O_X$ be any operator acting on the sites belonging to a region $X$. Suppose that we take a set of sites $\Omega\subseteq\Lambda$ which satisfies $l=\mathrm{dist}(X,\Lambda\setminus\Omega)$.
  Then,
  \begin{equation}
    \norm{(U_t^{H})^\dagger O_XU_t^{H}-(U_t^{H_\Omega})^\dagger O_XU_t^{H_\Omega}}\leq|X|\norm{O_X}\frac{(2\zeta_0|t|)^l}{l!},
  \end{equation}
  where $H_\Omega\coloneqq\sum_{Y\subseteq\Omega}h_Y$ is a Hamiltonian of the subsystem $\Omega$, $U_t^{H}=\exp(-itH)$ is a unitary time evolution operator, and $\zeta_0\coloneqq\max_{p\in\Lambda}\sum_{Z\ni p}|Z|\norm{h_Z}=\order{L^0}$ is the propagation velocity of the system which does not depend on the system size $L$.
  The norm $\|\cdot\|$ is the operator norm and $|X|$ denotes the number of sites in $X$.
\end{lem}
This lemma states that in many-body systems the time evolution of a local operator $O_X$ acting on a local region $X$ during time $T$ can be approximated by an operator acting on a set of sites within a distance of $\zeta_0T$ from $X$.

\subsection{Localizing the dissipators in the Redfield equation}
\label{sec:LRF}
Using the Lieb-Robinson bound, we can approximate the Redfield equation (5) so that its locality is apparent.
Focusing on the integrand on the right-hand side in Eq.~(\ref{eq:RedfieldSchr}), we find that the correlation function $C_{\mu\nu}(s)$ decays with the relaxation time $\tau_{\mathrm{B}}$ of $C_{\mu\nu}(s)$.
Thus, the range of the integral can be well approximated by $[0,T]$, where $T\sim\tau_{\mathrm{B}}$.
Therefore, it is legitimate to replace the upper bound of the integral with $T$.
We assume that $A_\mu$ acts on a local region $X_\mu$ and that the correlation function $C_{\mu\nu}(s)$ decays sufficiently fast as the distance between the two regions, $\mathrm{dist}(X_\mu,X_\nu)$, increases.

\begin{figure}
  \centering
  \includegraphics[width=8.6 cm]{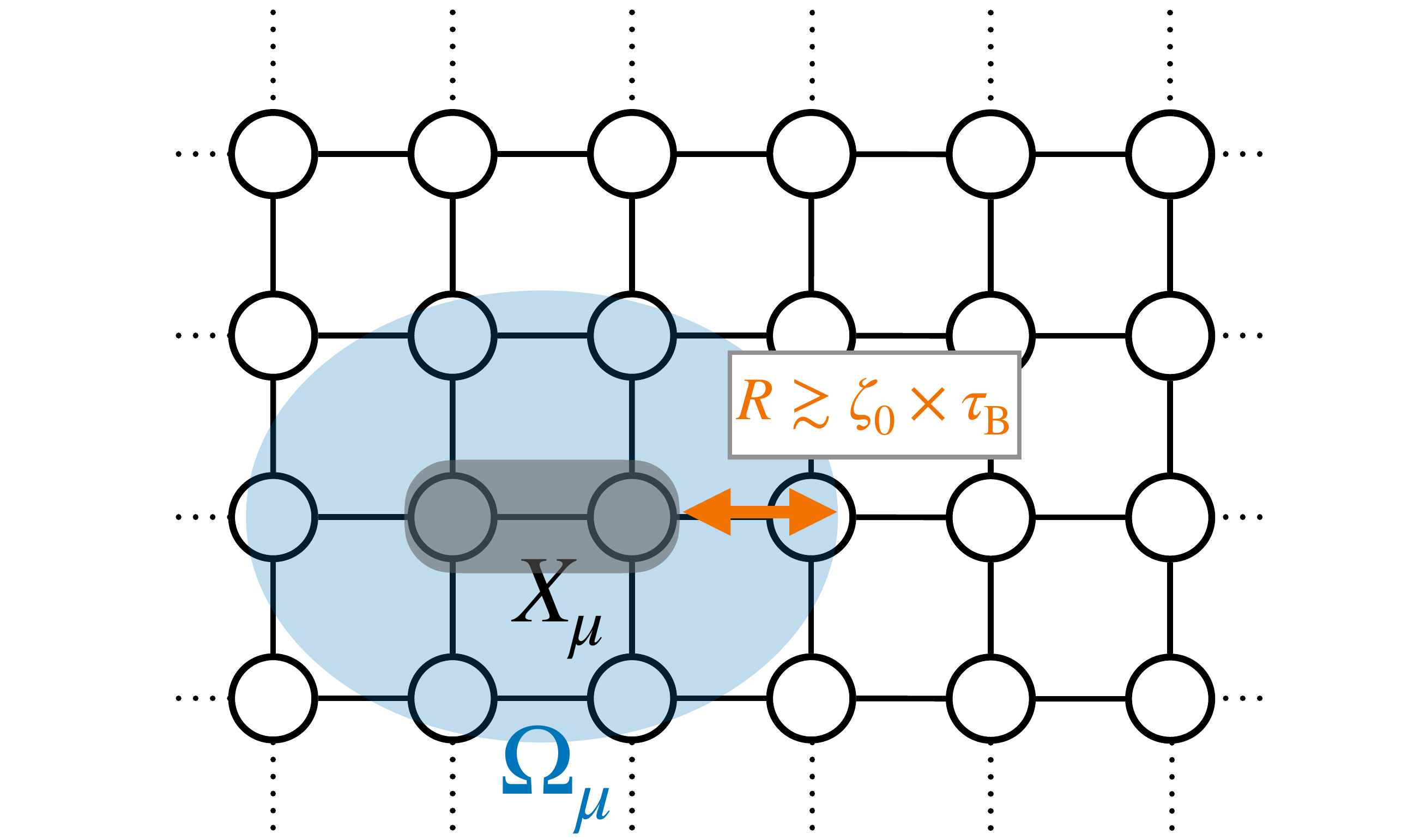}
  \caption{Subsystem $\Omega_\mu$ on which a local operator $A_\mu^\mathrm{loc}$ acts in the dissipator. Circles represent the sites of the system.
  The inner shaded region represents $X_\mu$, the set of sites on which $A_\mu$ acts.
  The outer shaded region represents the subsystem $\Omega_\mu$ on which the system-bath coupling $A_\mu\otimes B_\mu$ exerts during the relaxation time $\tau_\mathrm{B}$ of the bath.
  This subsystem is chosen to be composed of the sites within the distance shorter than $R\gtrsim\zeta_0\times\tau_{\mathrm{B}}$ from every site connected to the corresponding bath.}
  \label{fig:subsystem}
\end{figure}

Here, we can use the Lieb-Robinson bound in Lem.~\ref{lem:LRbound} and approximate $A_\mu(s)$ in Eq.~(\ref{eq:RedfieldSchr}) by
\begin{equation}
  A_\mu^{\mathrm{loc}}(s)\coloneqq\exp(isH_{\Omega_\mu})A_\mu\exp(-isH_{\Omega_\mu}),
\end{equation}
where ${\Omega_\mu}$ is the subsystem which is chosen to satisfy $\mathrm{dist}(X_\mu,\Lambda\setminus{\Omega_\mu}) \gtrsim \zeta_0\tau_{\mathrm{B}}$ (see Fig.~\ref{fig:subsystem}).
For example, if $X_\mu$ consists of a single site, $\Omega_\mu$ is constituted from sites whose distance from the single site is less than $R\gtrsim \zeta_0\tau_{\mathrm{B}}$.
Then, we obtain the local Redfield equation:
  \begin{equation}
    \label{eq:Redfieldloc}
    \begin{split}
      &\frac{d}{dt}\rho(t)=-i[H_\mathrm{S},\rho(t)]+\sum_{\mu,\nu}\int_0^\infty ds\left[C_{\mu\nu}(s)\right.\\
      &\left.\times(A_\nu^{\mathrm{loc}}(-s)\rho(t) A_\mu-A_\mu A_\nu^{\mathrm{loc}}(-s)\rho(t))+\mathrm{H.c.}\right],
    \end{split}
  \end{equation}
where all the operators $A_\nu^\mathrm{loc}$ and $A_\mu$ in the dissipator act on local subsystems.
When all $A_\mu$'s act on the same region $X$, the second term in Eq.~(\ref{eq:Redfieldloc}) is local, acting only on $\Omega$ which satisfies $\mathrm{dist}(X,\Lambda\setminus\Omega) \gtrsim \zeta_0\tau_{\mathrm{B}}$.
When $A_\mu$ acts on a spatially distant region $X_\mu$, the second term in Eq.~(\ref{eq:Redfieldloc}) is also local, if the correlation function $C_{\mu\nu}(s)$ of the bath decays sufficiently fast as the spatial distance $\mathrm{dist}(X_\mu,X_\nu)$ increases.


\subsection{Local Davies equation}
\label{sec:locDavies}
By making the rotating-wave approximation in Eq.~(\ref{eq:Redfieldloc}), the local Davies equation can be derived.
The local Davies equation has been used in the literature of quantum thermodynamics~\cite{Levy_2014,Barra2015,Hofer_2017,PhysRevE.97.022115,DeChiara_2018,RevModPhys.93.035008,Scali2021localmaster} because it can describe the relaxation to the Gibbs state at the temperature of a bath.

An operator $A_\mu$ can be decomposed into the sum of its frequency components $A^{\mathrm{loc}}_\mu(\omega)$ over $\omega$,
\begin{equation}
  \label{eq:FreqDecomploc}
  A_\mu=\sum_{\omega}A_\mu^\mathrm{loc}(\omega),
\end{equation}
where $A_\mu^{\mathrm{loc}}(\omega)$ is defined as
\begin{equation}
  A_\mu^{\mathrm{loc}}(\omega)\coloneqq \sum_{E_n^{\mathrm{loc}}-E_m^{\mathrm{loc}}=\omega}\ket{E_m^{\mathrm{loc}}}\bra{E_m^{\mathrm{loc}}}A_\mu\ket{E_n^{\mathrm{loc}}}\bra{E_n^{\mathrm{loc}}},
\end{equation}
with $E_n^{\mathrm{loc}}$ and $|E_n^{\mathrm{loc}}\rangle$ being an eigenvalue and the corresponding eigenstate of $H_{\Omega_\mu}$.
We note that the sum on the right-hand side runs over all pairs of energy levels in $H_{\Omega_\mu}$ whose spacing is equal to $\omega$.

By substituting  Eq.~(\ref{eq:FreqDecomploc}) into Eq.~(\ref{eq:Redfieldloc}), we obtain
\begin{widetext}
\begin{equation}
  \label{eq:locRedfieldFreq}
      \frac{d}{dt}\rho (t)=-i[H_\mathrm{S},\rho (t)]+\sum_{\mu,\nu}\sum_{\omega,\omega^\prime}\left[\left(\frac{1}{2}\gamma_{\mu\nu}(\omega)+i\eta_{\mu\nu}(\omega)\right)(A_{\nu}^{\mathrm{loc}}(\omega)\rho (t){A_\mu^{\mathrm{loc}}}^\dagger(\omega^\prime)-\rho (t){A_\mu^{\mathrm{loc}}}^\dagger(\omega^\prime)A_\nu^{\mathrm{loc}}(\omega))+\mathrm{H.c.}\right].
\end{equation}
\end{widetext}
We define the time scale $\tau_{\mathrm{S},\Omega_\mu}$ of the subsystem by a typical value of $|\omega-\omega^\prime|^{-1}$, where $\omega$ and $\omega^\prime$ are the energy-level spacings of $H_{\Omega_\mu}$.
By assuming that the time scale $\tau_{\mathrm{S},\Omega_\mu}$ is sufficiently shorter than 
the time scale $\tau_{\mathrm{SB}}$ of the system-bath interaction (see Eq.~(\ref{eq:tSB}) for its definition),
we can ignore the terms with $\omega\neq \omega^\prime$ in Eq.~(\ref{eq:locRedfieldFreq}), and obtain the local Davies equation as
\begin{equation}
    \label{eq:localRWA}
    \begin{split}
      \frac{d}{dt}\rho (t)=&-i[H_\mathrm{S}+H_\mathrm{LS},\rho (t)]\\
      &+\sum_{\mu,\nu}\sum_{\omega}\gamma_{\mu\nu}(\omega)\left(A_\nu^{\mathrm{loc}}(\omega)\rho (t){A_\mu^{\mathrm{loc}}}^\dagger(\omega)\right.\\
      &~~~~~\left.-\frac{1}{2}\left\{{A_\mu^{\mathrm{loc}}}^\dagger(\omega){A_\nu^{\mathrm{loc}}}(\omega),\rho (t)\right\}\right),
    \end{split}
\end{equation}
where the Lamb-shift Hamiltonian $H_\mathrm{LS}$ is defined as
\begin{equation}
  H_\mathrm{LS}\coloneqq\sum_{\mu,\nu}\sum_\omega\eta_{\mu\nu}(\omega){A_\mu^\mathrm{loc}}^\dagger(\omega)A_\nu^\mathrm{loc}(\omega).
\end{equation}
The Lindblad operators $A_\mu^{\mathrm{loc}}(\omega)$ in Eq.~(\ref{eq:localRWA}) induce the transitions between energy eigenstates of the Hamiltonian $H_{\Omega_\mu}$ of the local subsystem and satisfy the detailed balance condition for $H_{\Omega_\mu}$, provided that the bath is at thermal equilibrium.

In order for the rotating-wave approximation for the Hamiltonian of the local subsystem $H_{\Omega_\mu}$ to be valid,
the energy-level spacings of $H_{\Omega_\mu}$ is sufficiently large so that the time scale $\tau_{\mathrm{S},\Omega_\mu}$ of the subsystem defined by a typical value of $|\omega-\omega^\prime|^{-1}$ must be much smaller than the time scale $\tau_{\mathrm{SB}}$ of the system-bath interaction,
where $\omega$ and $\omega^\prime$ are the energy-level spacings of $H_{\Omega_\mu}$.
Since the energy-level spacings become smaller as the size of the subsystem becomes larger,
we cannot take the subsystem $\Omega_\mu$ too large.
In contrast, the replacement of $A_\mu(t)$ by $A_\mu^\mathrm{loc}(t)$ becomes a better approximation as we take a larger subsystem.
Therefore, we expect a trade-off relationship between the error arising from localization of the Redfield equation and the error arising from the rotating-wave approximation.
We numerically show in Sec.~\ref{sec:4} that such a trade-off relationship indeed exists.

\subsection{Local Nathan-Rudner equation}
\label{sec:LULE}
In a manner similar to the derivation of the NRE in Sec.~\ref{sec:ULE}, starting from Eq.~(\ref{eq:Redfieldloc}), we can obtain the local NRE where $A_\mu(\omega)$'s in Eq.~(\ref{eq:ULELindbladop}) are replaced by $A_\mu^{\mathrm{loc}}(\omega)$.
The Lindblad operators and the Lamb-shift Hamiltonian of the local NRE can be represented in terms of $A_\mu^\mathrm{loc}(\omega)$ as
\begin{align}
  \label{eq:locULELindop}
  L_\lambda&=\sum_\mu\sum_\omega \gamma^{1/2}_{\lambda\mu}(\omega)A_\mu^{\mathrm{loc}}(\omega),\\
  \label{eq:locULEHLS}
  \begin{split}
    H_\mathrm{LS}=&\sum_{\mu,\nu}\sum_{\omega,\omega^\prime}\left[\eta_{\mu\nu}\left(\frac{\omega+\omega^\prime}{2}\right)\right.\\
    &\left.+i\frac{\gamma_{\mu\nu}(\omega)-\gamma_{\mu\nu}(\omega^\prime)}{4}\right]{A_\mu^{\mathrm{loc}}}^\dagger(\omega)A_\nu^{\mathrm{loc}}(\omega^\prime).
  \end{split}
\end{align}
These Lindblad operators and the Lamb-shift Hamiltonian can be obtained from the diagonalization of $H_{\Omega_\mu}$, while it is necessary to diagonalize the full Hamiltonian $H_\mathrm{S}$ to calculate the Lindblad operators~(\ref{eq:ULELindbladop}) and the Lamb-shift Hamiltonian~(\ref{eq:ULEHLS}) of the original NRE.
The local NRE derived here is yet another efficient approach to avoiding the diagonalization of the full Hamiltonian in numerical calculations.

The Lindblad operator can also be written as
  \begin{equation}
    \label{eq:ULElocLindop}
    L_\lambda=\sum_\mu\int^\infty_{-\infty}ds g_{\lambda\mu}(-s)A_\mu^\mathrm{loc}(s).
  \end{equation}
The relaxation time of $g_{\mu\mu}(t)$ is also of the order of $\tau_{\mathrm{B}}$. 
Because we have chosen the subsystem $\Omega_\mu$ for $A_\mu$ so that $\mathrm{dist}(X_\mu,\Lambda\setminus\Omega_\mu)\gtrsim \zeta_0\tau_\mathrm{B}$, where $X_\mu$ is the support of $A_\mu$, $A_\mu(s)$ in Eq.~(\ref{eq:ULELindoptime}) can be replaced by $A_\mu^\mathrm{loc}(s)$ by using the Lieb-Robinson bound.
We find that Eq.~(\ref{eq:ULElocLindop}) reduces to Eq.~(\ref{eq:ULELindoptime}) in the limit of $\tau_\mathrm{B}\to0$. 
Since the error of the NRE relative to the Redfield equation is $\order{\tau_{\mathrm{B}}/\tau_{\mathrm{SB}}}$, the error of the local NRE from the Redfield equation vanishes in the limit of $\tau_{\mathrm{B}}\to0$.

\section{Numerical test of the local GKSL equations}
\label{sec:4}
In this section, we numerically test the validity of the local Davies equation and the local NRE derived in Sec.~\ref{sec:3}.
To test the validity, we compare the solutions of the local GKSL equations with those of the Redfield equation because the Redfield equation describes the dynamics most accurately among other QMEs in the situation where the Markov approximation is valid~\cite{PhysRevA.101.012103}.
Here, we show that the distance between the generator of the local GKSL equation and that of the Redfield equation becomes small if $\tau_{\mathrm{B}} \ll R/\zeta_0$, which is consistent with the condition shown in Table~\ref{tab:table1}.
This result guarantees that the dynamics in a sufficiently short time is described by the local GKSL equation with a small deviation from that of the Redfield equation.
However, it does not guarantee the correctness of the steady state because the small deviation of the generator can accumulate to grow exponentially in a sufficiently long time compared with $\tau_\mathrm{SB}$.
Even if the generators of time evolution change only slightly, the steady state can change significantly.
Therefore, we also investigate the deviation of the steady states of the local GKSL equations from those of the Redfield equation
to confirm that the steady states can also be accurately described by the local GKSL equations.

\subsection{Model}
We consider spinless fermions on a one-dimensional lattice with $L$ sites. The Hamiltonian of the system is given by
\begin{align}
  H_\mathrm{S}&=\omega_0\sum_{j=1}^L a_j^\dagger a_j -J\sum_{j=1}^{L-1}(a_j^\dagger a_{j+1}+a_{j+1}^\dagger a_j)\\
  &=:\sum_{i,j=1}^{L}h_{ij}a_i^\dagger a_j,
\end{align}
where $\omega_0$ is an on-site energy and $J>0$ is the hopping amplitude.
Here, $a_j$ and $a_j^\dagger$ represent the annihilation and creation operators at site $j$, and they satisfy the anticommutation relations $\{a_i,a_j^\dagger\}=\delta_{i,j}$.
The propagation velocity of the system is given by $\zeta_0=4J$ (see Lem.~\ref{lem:LRbound} in Sec.~\ref{sec:LRbound} for the definition of the propagation velocity).

\begin{figure}
  \centering
  \includegraphics[width=8.6 cm]{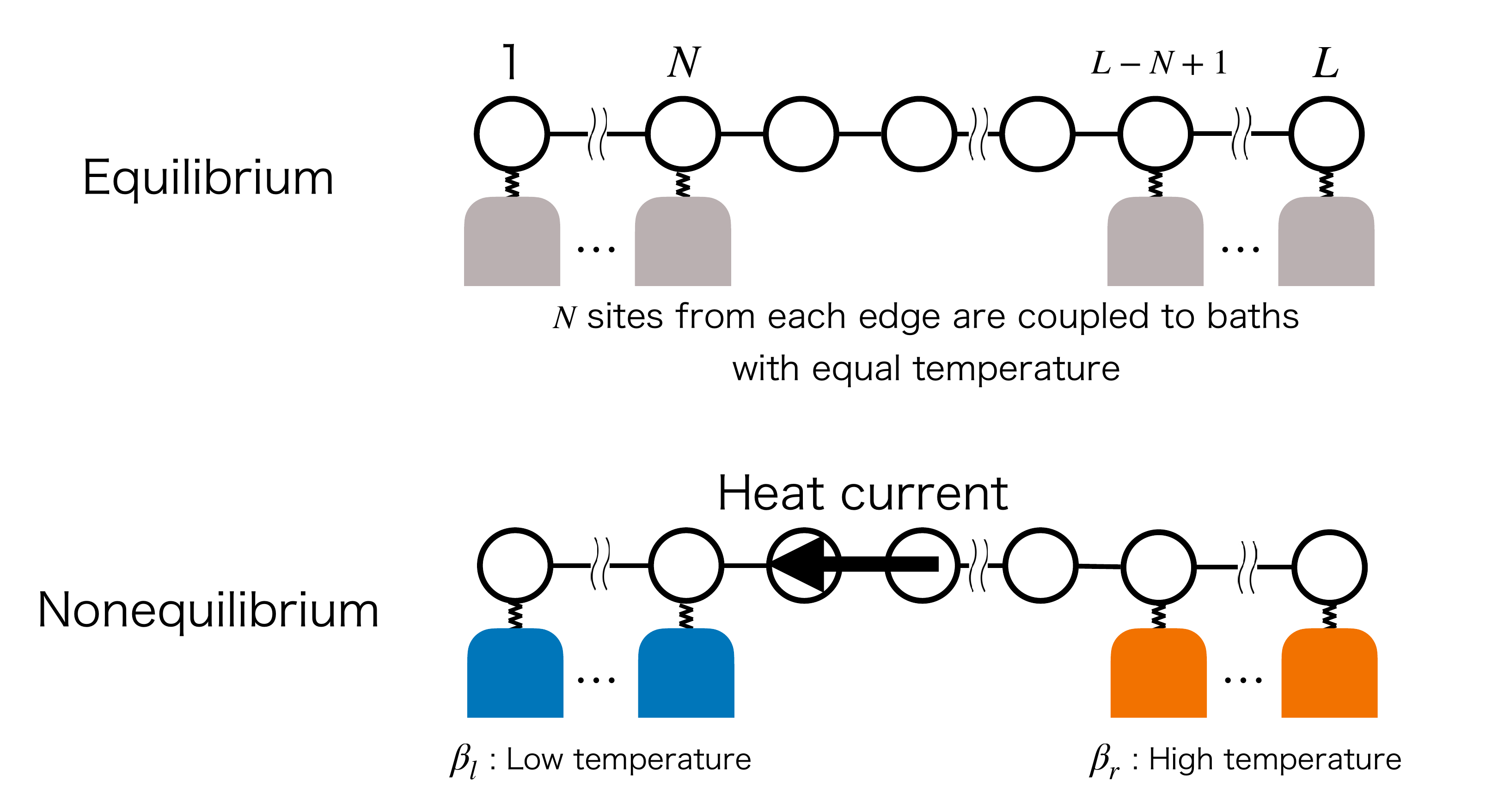}
  \caption{Schematic illustration of the model for numerical demonstration of the validity of the local GKSL equations.
  A spinless fermion chain is connected to baths so that an equilibrium (top, $\beta_l=\beta_r$) or a nonequilibrium (bottom, $\beta_l\neq\beta_r$) steady state is achieved.
  The round square under each site represents a bath connected to it.}
  \label{fig:setting}
\end{figure}

We couple $N$ sites from each edge of the one-dimensional lattice to identical baths constituted of free fermions (see Fig.~\ref{fig:setting}).
The state $\rho_{\mathrm{B},j}$ of the bath connected to site $j$ is assumed to be the Gibbs state at inverse temperature $\beta_j$ and the average number of fermions therein with wave number $\boldsymbol{k}$ is given by the Fermi-Dirac distribution $f_\beta(\omega(\boldsymbol{k}))=1/(1+e^{\beta_j\omega(\boldsymbol{k})})$, where $\omega(\boldsymbol{k})$ is the dispersion relation of fermions in the bath measured from the chemical potential.
Here, the chemical potentials of the baths are set to be equal.
The temperature $\beta_j$ is set to be $\beta_j=\beta_{l}~(\beta_r)$ if site $j$ is close to the left (right) edge.
In this section, we consider two settings: an equilibrium setting where $\beta_l=\beta_r$, and a nonequilibrium setting where there is a temperature difference, $\beta_l\neq\beta_r$ (see Fig.~\ref{fig:setting}).
The Hamiltonian $H_{\mathrm{B},j}$ of the bath at each site and the system-bath interaction Hamiltonian $H_{\mathrm{SB},j}$ at each site are given as
\begin{align}
  H_{\mathrm{B},j}&=\sum_{\boldsymbol{k}} \omega(\boldsymbol{k}) {c_{\boldsymbol{k}}^{(j)\dagger}} c_{\boldsymbol{k}}^{(j)},\\
  H_{\mathrm{SB},j}&=\frac{J_\mathrm{int}}{\sqrt{2\pi V}}\sum_{\boldsymbol{k}} (a_j^\dagger c_{\boldsymbol{k}}^{(j)}+{c_{\boldsymbol{k}}^{(j)\dagger}} a_j),
\end{align}
where $V$ is the volume of the baths, and $c_{\boldsymbol{k}}^{(j)}$ and $c_{\boldsymbol{k}}^{(j)\dagger}$ are the annihilation and creation operators of a fermion of the bath with wave number $\boldsymbol{k}$, respectively.
The total Hamiltonian is written as
\begin{equation}
  H_\mathrm{tot}=H_\mathrm{S}+\sum_{j=1}^{N}(H_{\mathrm{SB},j}+H_{\mathrm{B},j})+\sum_{j=L-N+1}^{L}(H_{\mathrm{SB},j}+H_{\mathrm{B},j}).
\end{equation}

To introduce the time scale $\tau_\mathrm{B}$ of the bath in this model, we assume that the density of states $D(\omega)$ of the baths in the limit $V\to\infty$ is well-approximated by the Cauchy-Lorentz distribution as
\begin{equation}
  D(\omega)\coloneqq\frac{1}{V}\sum_{\boldsymbol{k}}\delta(\omega-\omega(\boldsymbol{k}))\to \frac{2\tau_\mathrm{B}^{-1}}{{\tau_\mathrm{B}}^{-2}+(\omega-\omega_0)^2}.
\end{equation}
Here, we assume that the density of states peaks at $\omega_0$ for simplicity.
The interaction Hamiltonian at site $j$ can be rewritten as $H_{\mathrm{SB},j}=A_{1,j}\otimes B_{1,j}+A_{2,j}\otimes B_{2,j}$, where
\begin{equation}
  \begin{split}
  A_{1,j}\coloneqq a_j,&~A_{2,j}\coloneqq a_j^\dagger,\\
  B_{1,j}\coloneqq\frac{J_\mathrm{int}}{\sqrt{2\pi V}}\sum_{\boldsymbol{k}} {c_{\boldsymbol{k}}^{(j)\dagger}},&~B_{2,j}\coloneqq\frac{J_\mathrm{int}}{\sqrt{2\pi V}}\sum_{\boldsymbol{k}} c_{\boldsymbol{k}}^{(j)}.
  \end{split}
\end{equation}
Since the state of each bath is assumed to be given by the Gibbs state, the correlation functions $C_{\mu\nu}^{(j)}(t)=\tr[B_{\mu,j}^\dagger(t)B_{\nu,j}\rho_{\mathrm{B},j}]$ in the limit of $V\to\infty$ are written as
\begin{equation}
  \begin{split}
    C_{11}^{(j)}(t)&=J_\mathrm{int}^2\int_{-\infty}^{\infty}\frac{d\omega}{2\pi} e^{-i\omega t}(1-f_{\beta_j}(\omega))D(\omega),\\
    C_{22}^{(j)}(t)&=J_\mathrm{int}^2\int_{-\infty}^{\infty}\frac{d\omega}{2\pi} e^{i\omega t}f_{\beta_j}(\omega)D(\omega),\\
    C_{12}^{(j)}(t)&=C_{21}^{(j)}(t)=0,
  \end{split}
\end{equation}
and the power spectra are given as
\begin{equation}
  \begin{split}
    \gamma_{11}^{(j)}(\omega)&= J_\mathrm{int}^2(1-f_{\beta_j}(\omega))D(\omega),\\
    \gamma_{22}^{(j)}(\omega)&= J_\mathrm{int}^2 f_{\beta_j}(-\omega)D(-\omega),\\
    \gamma_{12}^{(j)}(\omega)&=\gamma_{21}^{(j)}(\omega)=0.
  \end{split}
\end{equation}
In the infinite-temperature limit $\beta_j=0$, the correlation functions are given by
\begin{equation}
  C_{11}^{(j)}(t)=C_{22}^{(j)}(t)=J_\mathrm{int}^2 e^{-t/\tau_\mathrm{B}}.
\end{equation}
Thus, at high temperature, $\tau_\mathrm{B}$ characterizes the time scale of decay of the correlation functions of the bath.

Here we derive the local GKSL equation for this model by using the method in Sec.~\ref{sec:3}.
To obtain the local Lindblad operators that describe the dissipation due to the bath at site $j$, we take the subsystem $\Omega_{R}^{(j)}$ composed of sites whose distance from site $j$ is shorter than $R$ (see Fig.~\ref{fig:subsystem}).
We call this $R$ the radius of the subsystem $\Omega_R^{(j)}$.
The Hamiltonian $H_{\mathrm{loc},j}$ of the subsystem is then given by
\begin{equation}
  \begin{split}
    H_{\mathrm{loc},j}=&~\omega_0\sum_{i\in\Omega_{R}^{(j)}} a_i^\dagger a_i\\
    &-J\sum_{\{i,i+1\}\subset\Omega_{R}^{(j)}}(a_i^\dagger a_{i+1}+a_{i+1}^\dagger a_i)\\
    =&:\sum_{i,l\in\Omega_{R}^{(j)}}h_{il}^{(\mathrm{loc},j)}a_i^\dagger a_l.
  \end{split}
\end{equation}
The matrix $h_{il}^{(\mathrm{loc},j)}$ can be diagonalized by an orthogonal matrix $O$ as
\begin{equation}
  \sum_{i,l}O_{mi}h_{il}^{(\mathrm{loc},j)}O_{nl}=\omega_m^{(\mathrm{loc},j)}\delta_{nm},
\end{equation}
and the subsystem Hamiltonian can be written as
\begin{equation}
  H_{\mathrm{loc},j}=\sum_{m=1}^{|\Omega_{R}^{(j)}|}\omega_m^{(\mathrm{loc},j)}d_m^\dagger d_m,
\end{equation}
where $d_m=\sum_{i\in\Omega_{R}^{(j)}}O_{mi}^*a_i$ is the annihilation operator of the energy eigenmode of the subsystem Hamiltonian.
Therefore, $A_{1,j}$ and $A_{2,j}$ can be decomposed as (see Eq.~(\ref{eq:FreqDecomploc}))
\begin{align}
  \label{eq:decomp1}
  A_{1,j}&=\sum_{m=1}^{|\Omega_{R}^{(j)}|}A_{1,j}^\mathrm{(loc)}(\omega_m^{(\mathrm{loc},j)})=\sum_{m=1}^{|\Omega_{R}^{(j)}|}O_{mj}d_m,\\
  \label{eq:decomp2}
  A_{2,j}&=\sum_{m=1}^{|\Omega_{R}^{(j)}|}A_{2,j}^\mathrm{(loc)}(-\omega_m^{(\mathrm{loc},j)})=\sum_{m=1}^{|\Omega_{R}^{(j)}|}O_{mj}^*d_m^\dagger.
\end{align}
We can use Eqs.~(\ref{eq:decomp1}) and (\ref{eq:decomp2}) to derive the local GKSL equation in a manner similar to the derivation in the previous section.

The obtained local GKSL equations and the Redfield equation can be written in the form of
\begin{equation}
  \label{eq:quadraticGKSL}
  \begin{split}
    \frac{d\rho}{dt}&=-i[H_\mathrm{S},\rho]+\sum_{m,n}\left(M_{mn}[w_m\rho,w_n]+\mathrm{H.c.}\right)\\
    &=:-i[H_\mathrm{S},\rho]+\mathcal{D}[\rho],
  \end{split}
\end{equation}
where $M=(M_{mn})$ is a Hermitian matrix and
\begin{equation}
  w_{2j-1}\coloneqq a_j+a_j^\dagger,~w_{2j}\coloneqq i(a_j-a_j^\dagger).
\end{equation}
Here, we neglect the small Lamb-shift Hamiltonian for the sake of simplicity.
For such QMEs, the steady-state expectation value of an observable which is written in the quadratic form of the annihilation and creation operators can be efficiently calculated~\cite{Prosen_2008,Prosen_2010,Prosen_2010PT,PhysRevA.95.052107}.
In addition, the modulus of the largest eigenvalue $\lambda_\mathrm{max}$ of $\mathcal{D}$ is also efficiently calculated from the matrix $M$ in Eq.~(\ref{eq:quadraticGKSL}) as $|\lambda_\mathrm{max}|=2|\tr M|$ as long as the dissipator has a single zero eigenvalue and the real part of all the other eigenvalues of the dissipator are negative (see Appendix~\ref{sec:appendixBB} for a proof).
This condition is satisfied in all the following numerical calculations.

\begin{figure}
  \centering
  \includegraphics[width=8.3 cm]{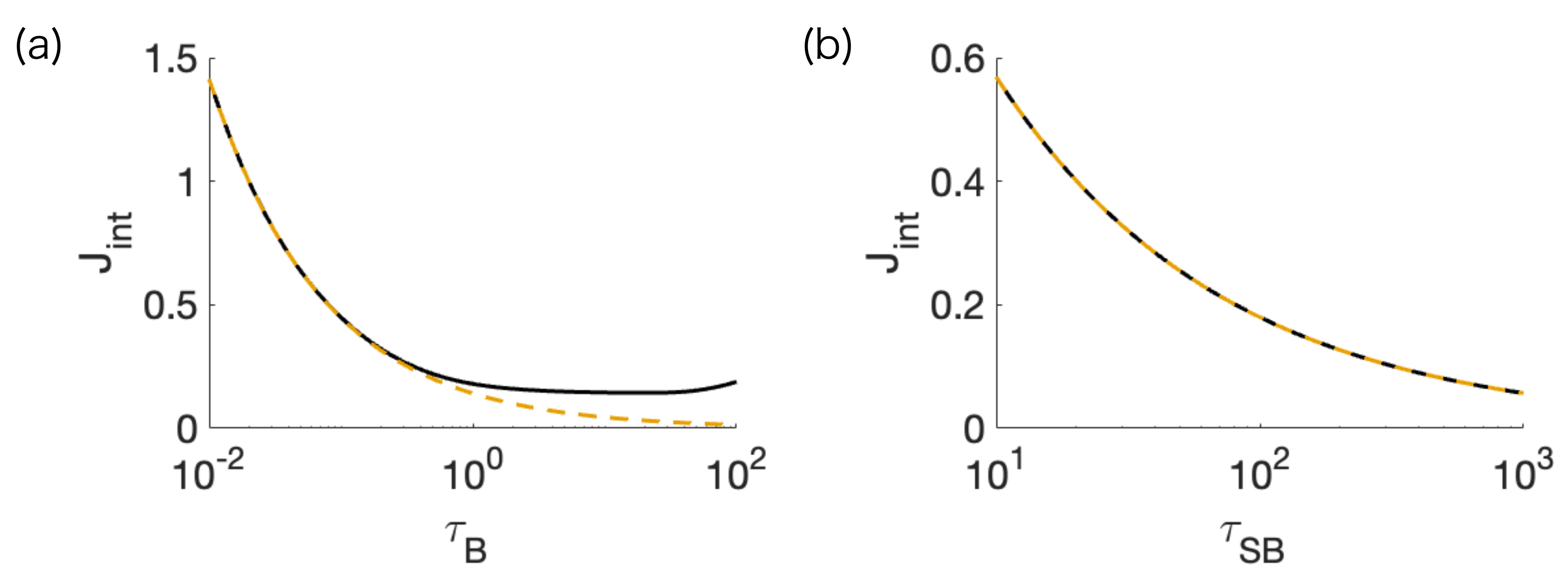}
  \caption{(a) The coupling $J_\mathrm{int}$ between the system and the bath plotted against $\tau_\mathrm{B}$ for fixed $\tau_\mathrm{SB}=100$.
  (b)) The coupling $J_\mathrm{int}$ plotted against $\tau_\mathrm{SB}$ for fixed $\tau_\mathrm{B}=1$.
  Both dashed curves represent $C(\tau_\mathrm{SB}\tau_\mathrm{B})^{-1/2}$ where $C$ is set to satisfy $J_\mathrm{int}=C(\tau_\mathrm{SB}\tau_\mathrm{B})^{-1/2}\simeq 0.18J$ for $\tau_\mathrm{B}=1$ and $\tau_\mathrm{SB}=100$.
  The parameters used are $L=128$, $N=16$, $J=1$, $\omega_0=0$, $\beta_l=0.5$ and $\beta_r=0.1$.}
  \label{fig:Jint}
\end{figure}

The norm of the dissipator $\mathcal{D}^{(\text{Redfield})}$ in the Redfield equation cannot be computed efficiently.
Therefore, although the norm $\|\mathcal{D}^{(\text{Redfield})}\|$ and the modulus $|\lambda_\mathrm{max}|$ of the largest eigenvalue do not generally coincide, here we expect the norm and the maximum eigenvalue to take similar values and characterize the time scale $\tau_\mathrm{SB}$ by $|\lambda_\mathrm{max}|$ as
\begin{equation}
  \tau_{\mathrm{SB}}^{-1}\coloneqq\|\mathcal{D}^{(\text{Redfield})}\|\simeq |\lambda_\mathrm{max}|= 2|\tr M^{(\text{Redfield})}|.
\end{equation}
This time scale $\tau_{\mathrm{SB}}$ can be adjusted by changing the coupling $J_\text{int}$ between the system and the bath.
We note that the coupling $J_\text{int}$ between the environment and the system and the time scales $\tau_\mathrm{B}$ and $\tau_\mathrm{SB}$ cannot be changed independently.
As shown in Fig.~\ref{fig:Jint}, the coupling $J_\mathrm{int}$ is proportional to $(\tau_\mathrm{SB}\tau_\mathrm{B})^{-1/2}$ for $\tau_\mathrm{B}\ll\tau_\mathrm{SB}$ and is almost constant for large $\tau_\mathrm{B}$.

In the following, we discuss the validity of the local GKSL equations by changing the ratios of the three time scales, $\tau_\mathrm{B}$, $\tau_\mathrm{SB}$, and $R\zeta_0^{-1}=R/4J$.
Since the condition $\tau_\mathrm{B}\ll\tau_\mathrm{SB}$ determines the validity of the Markov approximation, we choose $\tau_\mathrm{SB}$ as the parameter rather than $J_\mathrm{int}$ to ensure that the condition is satisfied.
Furthermore, as long as $\tau_\mathrm{SB}$ is set to be $10\tau_\mathrm{B}<\tau_\mathrm{SB}$ so that the Markov approximation is justified,
the deviations of the generator and the steady state of the local GKSL equations from those of the Redfield equation discussed in the following sections are almost independent of $\tau_\mathrm{SB}$.
Hence, in the following, we fix $\tau_\mathrm{SB}$ and consider the conditions under which the local GKSL equation is valid while varying $\tau_\mathrm{B}$ and $R$.
In the numerical calculation, the unit of time is set to be the inverse hopping rate $J^{-1}$.

\subsection{Deviation of the generator of the local GKSL equation from that of the Redfield equation}
\label{sec:distance}
To confirm the validity of the local GKSL equations derived in Sec.~\ref{sec:3}, we numerically calculate the distance between the generator of the time evolution of the Redfield equation and that of the derived local GKSL equation.
We fix $\tau_\mathrm{SB}$ and show that the distance becomes small under the condition $\tau_\mathrm{B}<R/\zeta_0$ by decreasing the value of $\tau_\mathrm{B}(\ll\tau_\mathrm{SB})$.
We define the distance between the generators of the two QMEs in terms of the ratio of the Hilbert-Schmidt norm of the difference of the matrix $M$ in Eq.~(\ref{eq:quadraticGKSL}) to that of $M$ of the Redfield equation as
\begin{equation}
  \label{eq:distance}
  \frac{\sqrt{\tr [|M^{\mathrm{(locGKSL)}}-M^{\mathrm{(Redfield)}}|^2]}}{\sqrt{\tr [|{M^{\mathrm{(Redfield)}}}|^2]}},
\end{equation}
where $|M|^2=M^\dagger M$.
This distance represents the error in the generator of the local GKSL equation.

\begin{figure}
  \centering
  \includegraphics[width=8.6 cm]{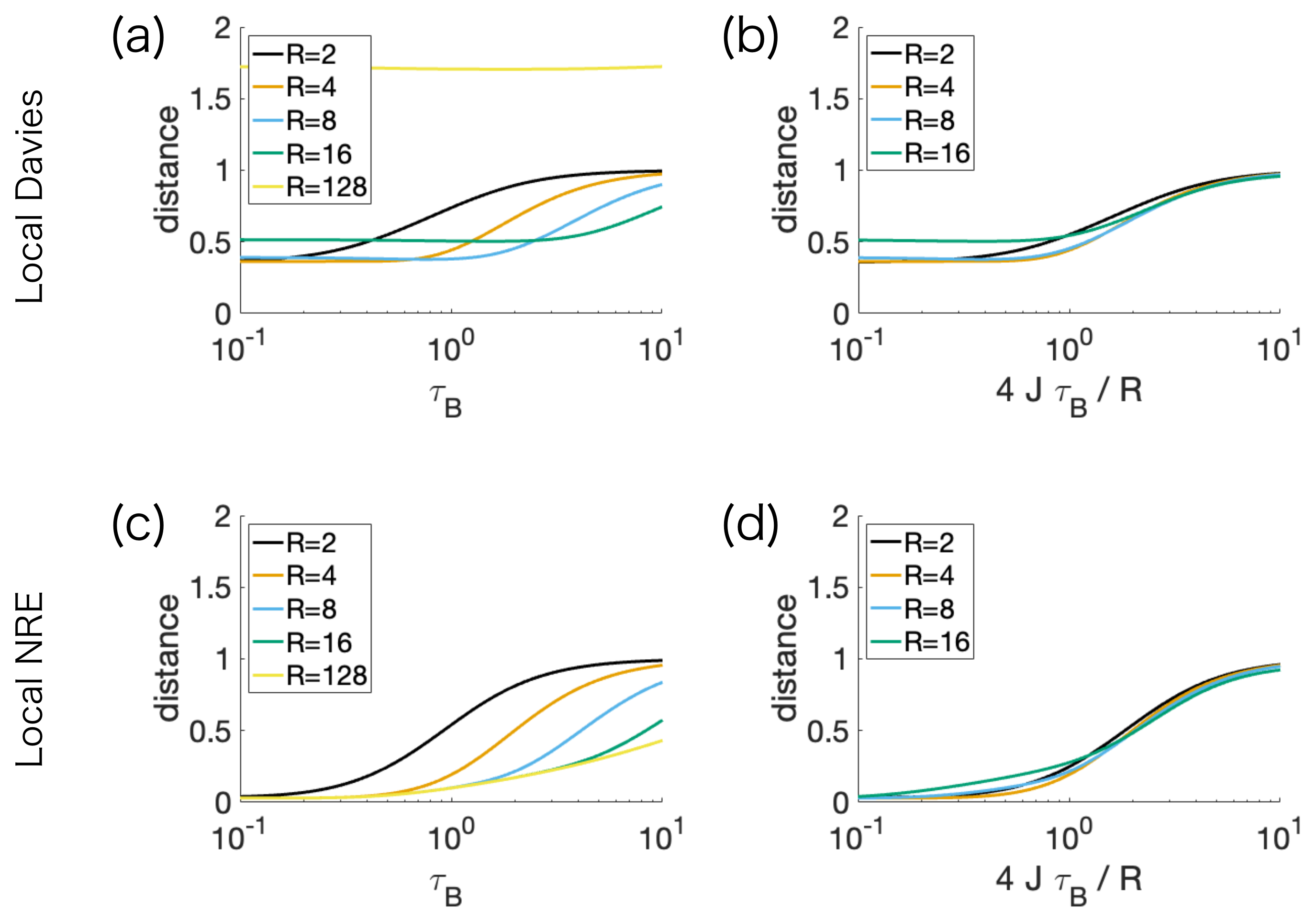}
  \caption{Distance between the generator of each local GKSL equation and that of the Redfield equation plotted against the relaxation time $\tau_\mathrm{B}$ of the baths.
  Here $R$ represents the radius of the local subsystem upon which the Lindblad operators act.
  The parameters used are $L=128$, $N=16$, $\tau_\mathrm{SB}=100$, $J=1$, $\omega_0=0$, $\beta_l=0.5$ and $\beta_r=0.1$.
  (a) Distance for the local Davies equation in the nonequilibrium setting. The distance varies nonmonotonically with increasing the radius.
  (b) The same quantity as in (a) with only the horizontal axis rescaled. The distance for the local Davies equation decreases significantly below $\tau_\mathrm{B}=R/4J$. 
  (c) Distance for the NRE in the nonequilibrium setting. The distance decreases with increasing the radius. 
  (d) The same quantity as in (c) with only the horizontal axis rescaled. The distance for the local NRE decreases significantly below $\tau_\mathrm{B}=R/4J$. 
  }
  \label{fig:diss}
\end{figure}

The distance (\ref{eq:distance}) depends on the radius $R$ of the subsystem chosen in the derivation of the local GKSL equation. 
For local GKSL equations with a different choice of the radius of the subsystem, the distances are plotted in Fig.~\ref{fig:diss} against the relaxation time $\tau_\mathrm{B}$ of the baths.
As local GKSL equations, we consider the local Davies equation (see Sec.~\ref{sec:locDavies}) and the local NRE (see Sec.~\ref{sec:LULE}).

The distance becomes small for both the local Davies equation and the local NRE when the relaxation time $\tau_\mathrm{B}$ is so short that the radius $R$ is larger than $\zeta_0\tau_\mathrm{B}$, where $\zeta_0=4J$ is the propagation velocity of the system.
This result is consistent with the condition used in the derivation of the local GKSL equation in Sec.~\ref{sec:3}, where the radius $R$ of the subsystem should be chosen to be larger than $\zeta_0\tau_B$ according to the Lieb-Robinson bound.
By comparing the two local GKSL equations, we find that the error is smaller for the local NRE when $\tau_\mathrm{B}$ is short, which indicates that the local NRE is a better approximation in terms of accuracy.
For the local Davies equation, the distance is nonzero in the limit of $\tau_\mathrm{B}\to0$ (Fig.~\ref{fig:diss}-(a,b)) while it vanishes for the local NRE (Fig.~\ref{fig:diss}-(c,d)) as mentioned in Sec.~\ref{sec:LULE}.
This is because the deviation due to the rotating-wave approximation, which is determined by the time scales $\tau_\mathrm{S,\Omega}$ and $\tau_\mathrm{SB}$, remains nonzero in the limit of $\tau_\mathrm{B}\to0$.
In addition, the distance for the local Davies equation does not decrease monotonically with increasing the radius, even though the distance for the local NRE decreases as the radius increases.
The origin of this behavior will be discussed in more detail in Sec.~\ref{sec:tradeoff}.

As shown in Appendix~\ref{sec:appendixB}, the behavior of the distance is quantitatively different depending on whether the radius $R$ is even or odd.
We emphasize that the difference is noticeable for large $\tau_\mathrm{B}$, where the localization of the Redfield equation is not justified in the first place.
Therefore, the difference does not affect the validity of the local GKSL equation.
In this section, we focus on only even $R$.

\subsection{Deviation of the steady state}
We numerically calculate the deviation in the steady state of the local GKSL equation from that of the Redfield equation.
We again fix $\tau_\mathrm{SB}$ and show that the deviation in the steady state also becomes small as long as the condition $\tau_\mathrm{B}<R/\zeta_0$ is satisfied.
The accuracy of the GKSL equation cannot be evaluated only by the distance between the generators of the local GKSL equation and the Redfield equation.
Nevertheless, we argue that the time coarse-grained dynamics or the steady state can accurately be described by the GKSL equation if the distance between the generators is small.
To demonstrate this, we evaluate the deviation of the steady states of the local GKSL equations from that of the Redfield equation.

For the steady state of each QME, we follow Refs.~\cite{Prosen_2008,Prosen_2010,Prosen_2010PT,PhysRevA.95.052107} to compute the $2L\times2L$-matrix $W=(W_{mn})$ defined by
\begin{equation}
  W_{mn}=\tr(w_m w_n \rho_{\mathrm{steady}}),
\end{equation}
where $\rho_{\mathrm{steady}}$ is the density matrix of the steady state.
The deviation of the steady state of the local GKSL equation from that of the Redfield equation is evaluated by the normalized maximal difference between the $W$ matrix of the steady state of the local GKSL equation and that of the Redfield equation:
\begin{equation}
  \Delta=\frac{\max_{mn}|W_{mn}^{\mathrm{locGKSL}}-W_{mn}^{\mathrm{Redfield}}|}{\max_{m\neq n}|W_{mn}^{\mathrm{Redfield}}|},
\end{equation}
where we normalize the deviation by the maximum off-diagonal element of the $W$ matrix of the Redfield equation.
We note that the diagonal elements of $W$ are always unity and do not reflect the state of the system.

\begin{figure}
  \centering
  \includegraphics[width=8.3 cm]{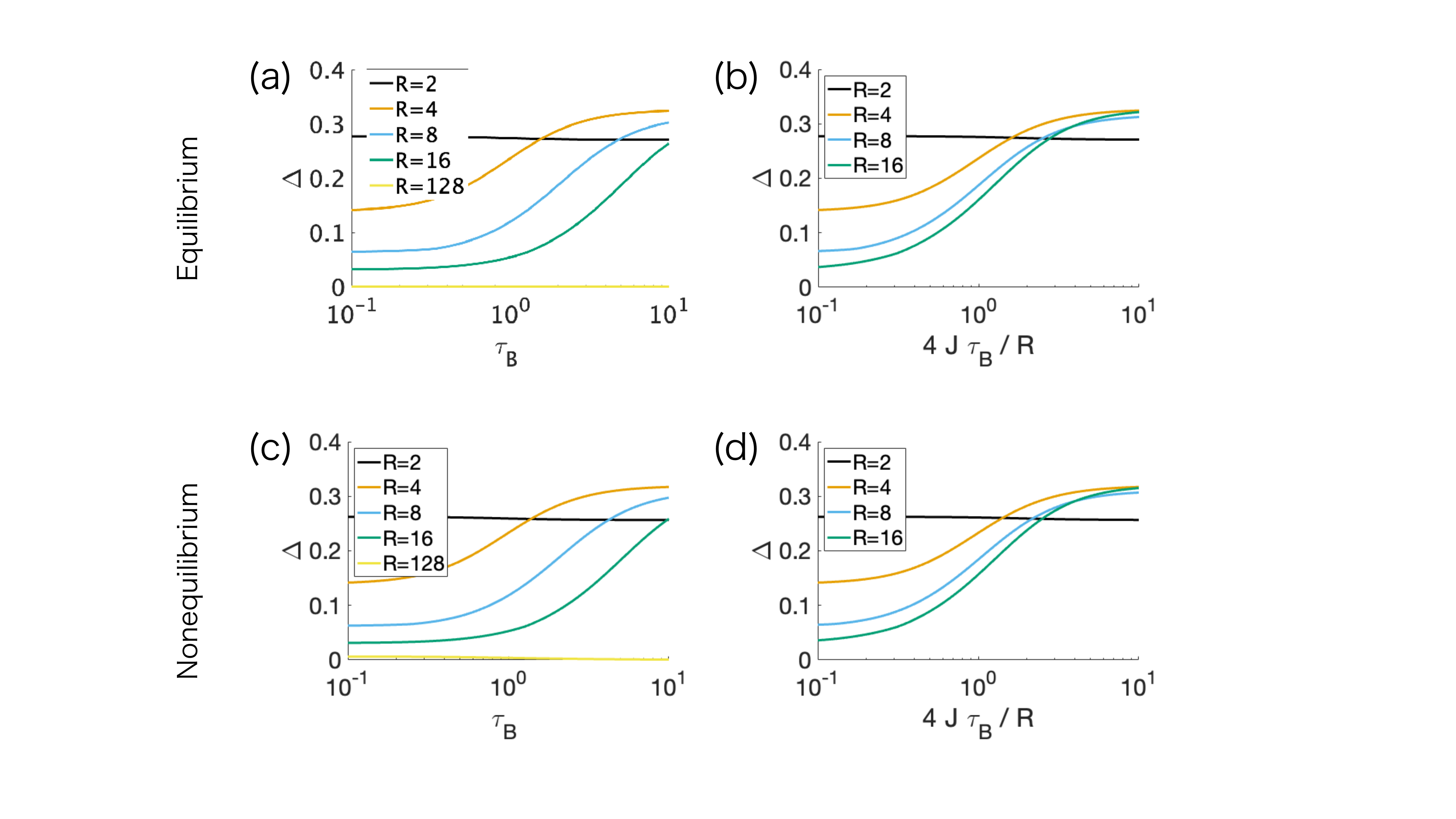}
  \caption{deviation of the steady state of the local Davies equation from that of the Redfield equation. In the local Davies equation, Lindblad operators act on local subsystems with radius $R$.
  The parameters are set to be $L=128$, $N=16$, $\tau_\mathrm{SB}=100$, $J=1$ and $\omega_0=0$.
  (a) Case of an equilibrium steady state, where the inverse temperatures of the baths are set to be equal: $\beta_l=\beta_r=0.1$.
  We note that the deviation between the Redfield equation and the Davies equation ($R=128$) vanishes because both lead to the same steady state (the Gibbs state).
  (b) The same quantity as in (a) with only the horizontal axis rescaled. 
  We see that the deviation significantly decreases around the region $4J\tau_\mathrm{B}/R<1$ where the radius $R$ of the subsystem is larger than $\zeta_0\tau_\mathrm{B}$ ($\zeta_0=4J$ is the propagation velocity) except for the case of $R=2$.
  (c) Case of a nonequilibrium steady state, where the inverse temperatures of the baths at the ends are $\beta_l=0.5$ for the left and $\beta_{r}=0.1$ for the right. The behavior of the deviation is qualitatively the same for the equilibrium and nonequilibrium cases except that the deviation for the Davies equation ($R=128$) is nonzero since the steady state is no longer the Gibbs state.
  (d) The same quantity as in (c) with only the horizontal axis rescaled. 
  As in (b), we also see that the deviation significantly decreases around the region $4J\tau_\mathrm{B}/R<1$ where the radius $R$ of the subsystem is larger than $\zeta_0\tau_\mathrm{B}$ except for the case of $R=2$.}
  \label{fig:LRWA}
\end{figure}

\begin{figure}
  \centering
  \includegraphics[width=8.6 cm]{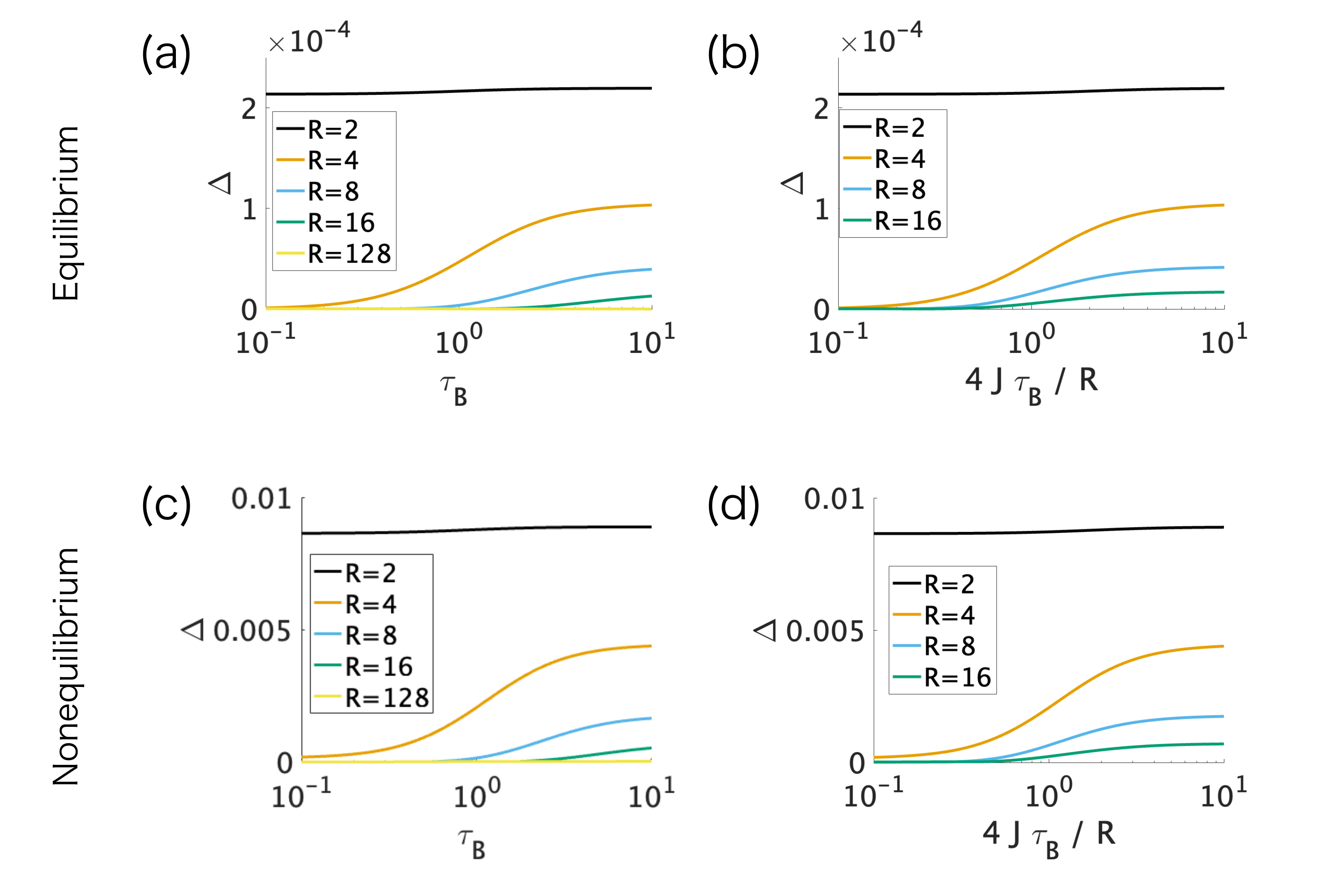}
  \caption{deviation of the steady state described by the local NRE from that of the Redfield equation. In the local NRE, Lindblad operators act on local subsystems with the radius $R$.
  The parameters are set to be $L=128$, $N=16$, $\tau_\mathrm{SB}=100$, $J=1$ and $\omega_0=0$.
  (a) Case of an equilibrium steady state, where the inverse temperatures of the baths are set to be equal as $\beta_l=\beta_r=0.1$. There is a nonzero deviation in the steady state of the $R=2$ local NRE even in the limit of $\tau_\mathrm{B}\to0$, where the local NRE reduces to the NRE.
  (b) The same quantity as in (a) with only the horizontal axis rescaled.
  We see that the deviation becomes small in the region $4J\tau_\mathrm{B}/R<1$.
  (c) Case of a nonequilibrium steady state, where the inverse temperatures of the baths are $\beta_l=0.5$ and $\beta_{r}=0.1$.
  deviations are larger than those in the equilibrium case.
  (d) The same quantity as in (c) with only the horizontal axis rescaled.
  We see that the local NRE can describe the steady state with a sufficiently small deviation in the region $4J\tau_\mathrm{B}/R<1$.}
  \label{fig:LULE}
\end{figure}

The deviations for each local GKSL equation are evaluated for both an equilibrium steady state and a nonequilibrium steady state.
The results for the local Davies equation and the local NRE are shown in Figs.~\ref{fig:LRWA} and \ref{fig:LULE}, respectively.
The deviations of the steady states for both of the local GKSL equations become small when $\tau_\mathrm{B}<R/4J$ in a similar way as the distance between the generators.
This is consistent with the condition for applying the approximation to the Redfield equation using the Lieb-Robinson bound described in Sec.~\ref{sec:3}.

For the local Davies equation (Fig.~\ref{fig:LRWA}), the deviations of the steady states are reduced to about 10\% for $R\geq 4$ and $\tau_{\mathrm{B}} \ll R/\zeta_0$, even though the generators produce deviations larger than 40\% (see Fig.~\ref{fig:diss} (b)).
In the equilibrium case, the Gibbs state is the exact steady state of the Davies equation and the steady state described by the local Davies equation approaches the Gibbs state as $R$ increases despite the large deviation in the generators.
The deviation in the nonequilibrium steady state of the local Davies equation is also as small as that in the equilibrium steady state even though the nonequilibrium steady state of the Davies equation and that of the Redfield equation do not coincide.
The deviation is nonzero in the limit of $\tau_\mathrm{B}\to0$ since the generator of the local Davies equation has nonzero deviation from that of the Redfield equation as mentioned in Sec.~\ref{sec:distance}.

For the local NRE (Fig.~\ref{fig:LULE}), the deviation is smaller than that of the local Davies equation in both equilibrium and nonequilibrium cases.
In contrast to the local Davies equation, the deviation in the steady state tends to be larger in the nonequilibrium case than the equilibrium case (Fig.~\ref{fig:LULE} (c)).
For $R=2$ and $4$, the steady state of the local NRE has nonzero deviations even in the limit of $\tau_\mathrm{B}\to0$, whereas the steady states for the local NREs of $R\geq 8$ and the NRE ($R=L=128$) have negligible deviations in the same limit.
This result can be attributed to the fact that small deviations in the generators of the dynamics may accumulate in the long time and result in a nonzero deviation of the steady state.
In fact, we have numerically confirmed that the distance of the generator vanishes in the limit of $\tau_\mathrm{B}\to0$.
There may be such deviations even for $R>4$ although they are too small to be seen and buried in numerical errors.

\begin{figure}
  \centering
  \includegraphics[width=8.6 cm]{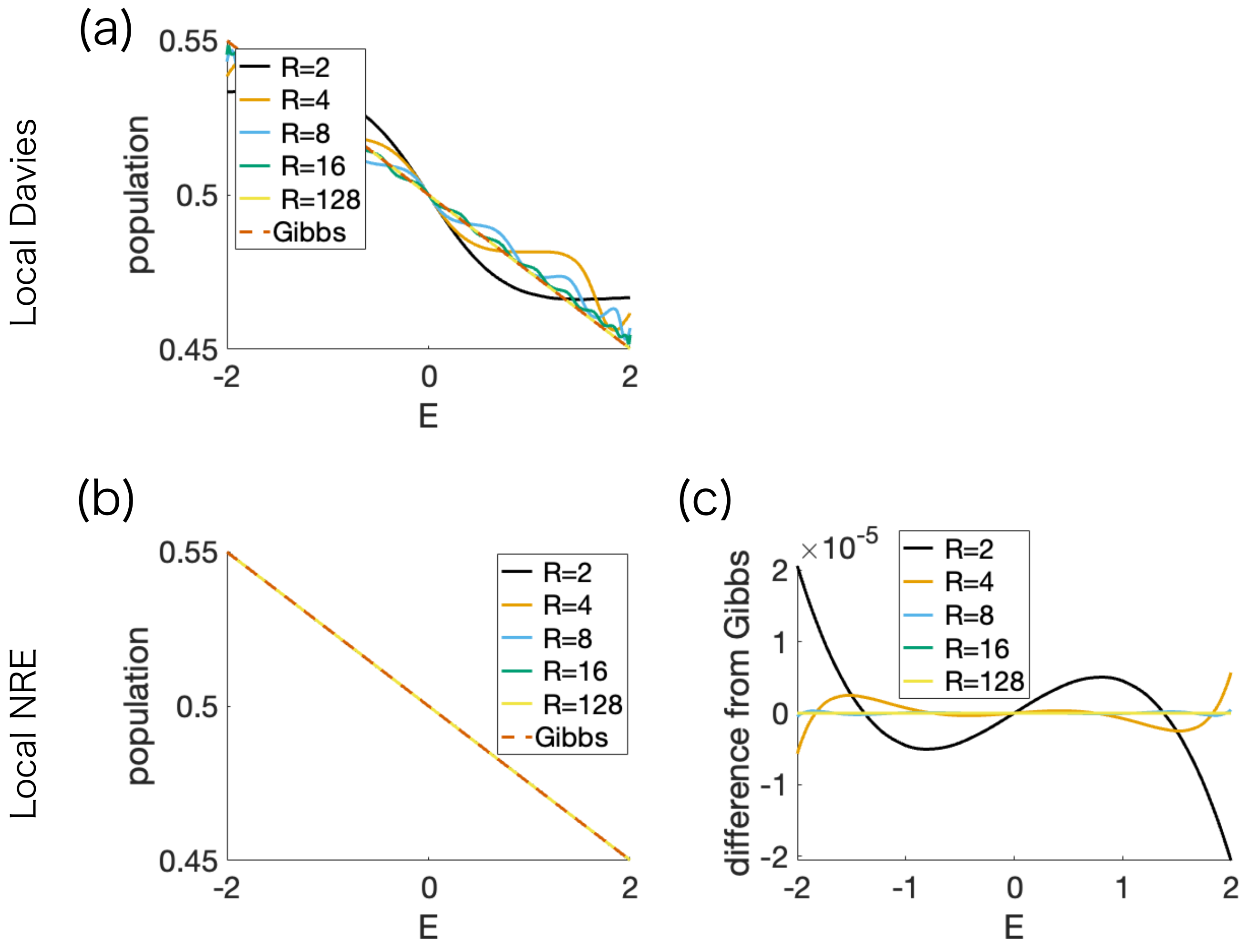}
  \caption{Population of the energy eigenmodes in the steady states of the local GKSL equations and that in the Gibbs state.
  The paremeters are set to be $L=128$, $N=16$, $\tau_\mathrm{SB}=100$, $\tau_\mathrm{B}=1$, $\beta_l=\beta_r=0.1$, $J=1$ and $\omega_0=0$.
  (a) Populations of the energy eigenmodes $\langle a_m^\dagger a_m\rangle~(\omega_m=E)$ in the steady states of the local Davies equations and the Gibbs state are plotted against the energy $E$.
  (b) Populations of the energy eigenmodes in the steady states of the local NRE and the Gibbs state are plotted against energy $E$.
  (c) Difference between the populations in the steady states of the local NRE and those in the Gibbs state are plotted against energy.}
  \label{fig:popilations}
\end{figure}

In additon to the deviation of the steady states, we show the populations of energy eigenmodes in the steady states of the local GKSL equations to see how well the local GKSL equations can describe thermalization.
The energy eigenmodes are obtained by diagonalizing the Hamiltonian of the system as $H_\mathrm{S}=\sum_m \omega_m a_m^\dagger a_m$.
In Fig.~\ref{fig:popilations}, the populations $\langle a_m^\dagger a_m\rangle$ in the steady states of the local Davies equation and the local NRE are plotted against the corresponding eigenenergy $E=\omega_m$.
To compare them with those at thermal equilibrium, we also show the populations in the Gibbs state $\rho=\exp(-\beta H_\mathrm{S})/\mathrm{tr}[\exp(-\beta H_{\mathrm{S}})]$.

It can be seen that both the local Davies equation and the local NRE accurately describe the population at thermal equilibrium within the same order of the deviation $\Delta$ in the steady state.
Therefore, as in the case of the deviation $\Delta$, it can be seen that the local NRE more accurately describes thermal equilibration than the local Davies equation.
Furthermore, in both cases, as the radius $R$ increases, the population in the steady state approaches that in the thermal equilibrium state.
We also find that the population in the steady state of the local Davies equation has the shape of $R$-step stairs.

\subsection{Trade-off between two deviations in the steady state of the local Davies equation}
\label{sec:tradeoff}
\begin{figure}
  \centering
  \includegraphics[width=8.3 cm]{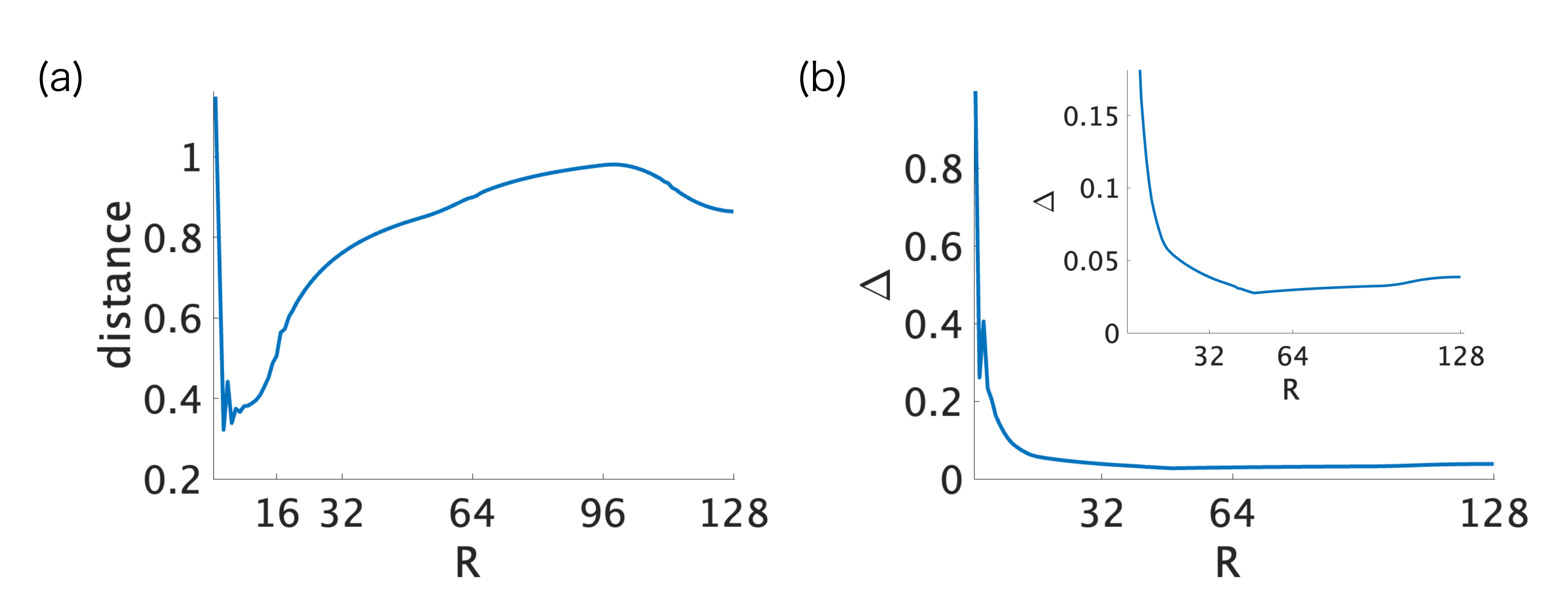}
  \caption{deviations of the generator (a) and the steady state (b) of the local Davies equation from those of the Redfield equation plotted against the radius $R$ of the local subsystem.
  The parameters are set to be $L=128$, $N=16$, $\tau_\mathrm{SB}=10$, $\tau_\mathrm{B}=1$, $\beta_l=0,5$, $\beta_r=0.1$, $J=1$ and $\omega_0=0$.
  As $R$ increase, both deviations initially decrease and then begin to increase at certain values of $R$. For (b), this can be clearly seen in the inset.}
  \label{fig:tradeoff}
\end{figure}

  We numerically show a trade-off relationship between two deviations from the Redfield equation caused by the rotating-wave approximation and the localization using the Lieb-Robinson bound.
  We fix $\tau_\mathrm{SB}$ and $\tau_\mathrm{B}$ and change the radius $R$ of the subsystem to investigate how the deviations in the generator and the steady state change as the time scale $\tau_{\mathrm{S,\Omega}}$ of the subsystem is changed.
  As shown in Fig.~\ref{fig:tradeoff}, both the deviation in the generator and the deviation in the steady state decrease with increasing $R$ because the deviation due to localization becomes smaller.
  However, the deviations begin to increase at certain values of $R$ because the deviation caused by the rotating-wave approximation becomes larger.
  The deviation in the generator and that in the steady state behave differently. While the deviation in the generator increases rapidly after $R=16$, the deviation in the steady state first decrease and then gradually increases from $R=32$.
  This result shows that when one uses the local Davies equation, the size of the subsystem should be taken appropriately depending on whether one focuses on the dynamics or the steady state.

\section{Discussion}
\label{sec:5}
We discuss the advantages of the local GKSL equations from the viewpoint of the numerical analysis of open quantum many-body systems compared with the existing GKSL equations.
To numerically simulate the dynamics of an open quantum many-body system, for practical purposes, it is desirable that the QME satisfies the following three conditions:
(i) The QME preserves the positive semidefiniteness of the density matrix, i.e., it takes a GKSL form as Eq.~(\ref{eq:GKSL}). As a consequence, the QME can be efficiently solved by the Monte Carlo method~\cite{breuer2002theory, AJDaley}.
(ii) The number of Lindblad operators of the QME is small enough to be numerically tractable, since we have to calculate the probabilities of quantum jumps caused by the Lindblad operators at each iteration of the Monte Carlo method~\cite{AJDaley}.
For example, in cases where the number of Lindblad operators grows exponentially with respect to the system size, it is necessary to compute the probabilities associated with an exponential number of quantum jumps at each iteration of the Monte Carlo method.
This exponential growth requires a large numerical cost for a large system size.
(iii) The numerical cost for the computation of the Lindblad operators in Eqs.~(\ref{eq:RWALind}) and (\ref{eq:ULELindbladop}) is not so high.
For example, it is practically impossible to compute Lindblad operators of the Davies equation and the NRE in cases where the diagonalization of the Hamiltonian is not feasible, such as in many-body systems.

We first discuss whether or not each of these three conditions holds in the existing microscopically derived GKSL equations introduced in Sec. II.
All GKSL equations satisfy condition (i). However, the Davies equation does not satisfy conditions (ii) and (iii).
The Lindblad operators in Eq.~(\ref{eq:RWALind}) represent the transitions between energy levels.
Therefore, there are as many Lindblad operators as possible energy transitions due to interactions with the bath, the number of which is exponentially large in many-body systems.
In addition, to obtain Lindblad operators, we need to diagonalize the Hamiltonian $H_\mathrm{S}$ of the many-body system of interest, find the eigenvalue $E_n$ and its eigenstate $\ket{E_n}$, and calculate $A_\mu(\omega)$.
On the other hand, the NRE satisfies the condition (ii)
because the number of Lindblad operators is not more than the number of interaction terms in Eq.~(\ref{eq:SBint}).
On the condition (iii) for the NRE, diagonalization of $H_S$ is necessary to obtain the exact Lindblad operators of the NRE.

In contrast, the local GKSL equations satisfy all these three conditions (i)-(iii).
Our method based on the Lieb-Robinson bound, which is universally present in many-body systems, is applicable to general many-body systems.
Owing to the locality, the number of Lindblad operators representing the effect from a bath is at most limited to the dimension of the space of those operators that act on the local subsystem, and the cost of computing the Lindblad operators is as small as the cost of diagonalizing the operators acting on the local subspace.
We note that Ref.~\cite{PhysRevB.102.115109} also proposes a method to avoid diagonalization of the Hamiltonian in the computation of the Lindblad operator of the NRE by truncating the expansion of $e^{itH}Xe^{-itH}$ as $e^{itH}Xe^{-itH}\simeq\sum_{n=0}^{N}(it)^n(\mathrm{ad}H)^nX/n!$, and it is actually applied in Ref.~\cite{PhysRevB.109.L180408}.

These three conditions (i)-(iii) make it easier to compute the dynamics of many-body systems described by the local GKSL equations. The computational cost required for computing the dynamics of the system by the Monte Carlo method is given by $M\times D^2$, where $D$ is the dimension of the Hilbert space of the system and $M$ is the number of sampled quantum trajectories~\cite{AJDaley}.
Here, $M$ is smaller than $D$ for most cases~\cite{AJDaley}.
In the case of the GKSL equation which requires the diagonalization of the Hamiltonian of the entire system, the computational cost of diagonalization is of the order of $D^3$.
Thus, the total computational cost for computing the dynamics with such a GKSL equation is also of the order of $D^3$.
On the other hand, in the case of the local GKSL equations based on the Lieb-Robinson bound, diagonalization of the Hamiltonian of the local subsystem is required to compute the Lindblad operators, and the computational cost stays roughly the same with respect to the system size since the dimension of the local subsystem is independent from the system size.
Therefore, the total computational cost for computing the dynamics with the local GKSL equations is of the order of $M\times D^2$.
Thus, the use of the local GKSL equations reduces the computational cost in the numerical analysis of open quantum many-body systems.

\section{Conclusion}
\label{sec:6}
We have proposed a method of microscopic derivation of the local GKSL equation by combining the localization of the Redfield equation on the basis of the Lieb-Robinson bound and the existing derivations of the GKSL equation~\cite{Davies1974,Cohen-Tannoudji_1986,LIDAR200135,PhysRevA.78.022106,PhysRevA.79.032110,PhysRevA.88.012103,Mozgunov2020completelypositive,PhysRevB.97.035432,PhysRevB.102.115109,Davidovic2020completelypositive,doi:10.1063/1.4907370,https://doi.org/10.48550/arxiv.1710.09939,PhysRevA.100.012107,PhysRevE.104.014110,PhysRevE.76.031115,PhysRevA.102.032207,Trushechkin_2016}.
While other derivations of the local GKSL equation~\cite{PhysRevE.76.031115,PhysRevA.102.032207,schnell2023global,Trushechkin_2016} treat the couplings between sites perturbatively, the causal structure given by the Lieb-Robinson bound makes the derivation valid for more general cases.
This derivation shows that the locality of the Lindblad operators is determined by the relaxation time $\tau_\mathrm{B}$ of the bath and the propagation velocity $\zeta_0$ of the system.
In this paper, we have introduced the local Davies equation and the local NRE; the former is derived from the local Redfield equation with the rotating-wave approximation, and the latter is derived from the local Redfield equation with the derivation of the NRE.
The different local GKSL equations can be used depending on the purpose of applications.
The local Davies equation can be used for thermodynamic settings, such as the case where dissipation due to a bath satisfies the detailed-balance condition for each subsystem.
The local NRE can be used to accurately describe a steady state as well as the dynamics.

By the numerical calculations for the one-dimensional tight-binding fermion chain, we have shown that the deviations of the generators and the steady state of the local GKSL equations from those of the Redfield equation become small when localization of the Redfield equation is performed so as to be compatible with the Lieb-Robinson bound, i.e., the radius of the local subsystem on which Lindblad operators act is taken to be larger than $\zeta_0\tau_\mathrm{B}$.
These numerical results support the validity of our derivation.
However, since small deviations in the generators may accumulate over a long time, a small deviation in the generators does not necessarily guarantee a small deviation in the steady state.
While we have numerically shown that the deviations in the steady state are indeed small for some cases, it is still an open problem how the behavior of deviations in a steady state can be understood from the microscopic derivation.

The local GKSL equation is a QME that can efficiently analyze open quantum many-body systems.
However, the local GKSL equation previously discussed in literature~\cite{RevModPhys.94.045006,PhysRevA.107.062216} has been given phenomenologically and does not reflect details of baths such as relaxation times.
Our microscopic derivation allows us to efficiently analyze open quantum many-body systems in such a manner as to accommodate detailed properties of surrounding baths.

\section*{Acknowledgement}
K.S. was supported by KAKENHI Grant No.~JP23KJ0730 from the Japan Society for the Promotion of Science (JSPS) and FoPM, a World-leading Innovative Graduate Study Program, the University of Tokyo.
M.N. was supported by KAKENHI Grant No.~JP20K14383 and No.~JP24K16989 from the JSPS.
T.M. was supported by KAKENHI Grant No.~JP21H05185 from the JSPS and PRESTO Grant No.~JPMJPR2259 from the Japan Science and Technology Agency (JST).
M.U. was supported by KAKENHI Grant No.~JP22H01152 from the JSPS.
We gratefully acknowledge the support from the CREST program ``Quantum Frontiers" (Grant No. JPMJCR23I1) by the Japan Science and Technology Agency.
\onecolumngrid
\appendix

\section{Derivation of the Nathan-Rudner equation in the frequency domain}
\label{sec:appendixA}
For the sake of self-containedness, we provide a derivation of the Nathan-Rudner equation (NRE)~\cite{PhysRevB.102.115109}.
While the original derivation in Ref.~\cite{PhysRevB.102.115109} is performed in the time domain, here we present another derivation in the frequency domain.
In the derivation of the NRE, we only assume that the time scale $\tau_\mathrm{SB}$ of the time evolution caused by the bath is much larger than the relaxation time $\tau_\mathrm{B}$ of the bath i.e., $\tau_{\mathrm{SB}}\gg\tau_\mathrm{B}$,
--- this condition is assumed to derive the Redfield equation.
Therefore, the NRE can describe the dynamics of an open system within the error of the same order as that of the Redfield equation~\cite{PhysRevB.102.115109}.

Let us begin with Eq.~(\ref{eq:frequencyMRSchr}).
If the condition $\tau_\mathrm{SB}\gg\tau_\mathrm{B}$ is satisfied, there always exists an intermediate time scale $\Delta t$ such that $\tau_\mathrm{B}\ll\Delta t\ll\tau_\mathrm{SB}$.
To derive the NRE, we modify the terms in Eq.~(\ref{eq:frequencyMRSchr}) with $|\omega-\omega^\prime|>\omega_c=\Delta t^{-1}$ so that the dynamics of the obtained GKSL equation does not deviate from that of the Redfield equation in the coarse-grained time scale larger than $\Delta t$.
The idea of the derivation presented here is fundamentally the same as that in Ref.~\cite{Davidovic2020completelypositive}, with a slightly different definition of the Lamb-shift Hamiltonian.

Expanding the Lamb-shift term in  Eq.~(\ref{eq:frequencyMRSchr}), we obtain
\begin{equation}
  \label{eq:beforeULE}
  \begin{split}
  \frac{d}{dt}\rho=&-i[H_\mathrm{S},\rho ]+\sum_{\mu,\nu}\sum_{\omega,\omega^\prime}\left[\frac{\gamma_{\mu\nu}(\omega)+\gamma_{\mu\nu}(\omega^\prime)}{2}+i(\eta_{\mu\nu}(\omega)-\eta_{\mu\nu}(\omega^\prime))\right]\left(A_\nu(\omega)\rho A_\mu^\dagger(\omega^\prime)-\frac{1}{2}\{A_\mu^\dagger(\omega^\prime)A_\nu(\omega),\rho \}\right)\\
  &+i\sum_{\mu,\nu}\sum_{\omega,\omega^\prime}\frac{1}{2}\left(\eta_{\mu\nu}(\omega^\prime)+i\frac{\gamma_{\mu\nu}(\omega)-\gamma_{\mu\nu}(\omega^\prime)}{2}\right)\left(A_\mu^\dagger(\omega^\prime)A_\nu(\omega)\rho -A_\nu(\omega)\rho A_\mu^\dagger(\omega^\prime)\right)\\
  &+i\sum_{\mu,\nu}\sum_{\omega,\omega^\prime}\frac{1}{2}\left(\eta_{\mu\nu}(\omega)+i\frac{\gamma_{\mu\nu}(\omega)-\gamma_{\mu\nu}(\omega^\prime)}{2}\right)\left(A_\nu(\omega)\rho A_\mu^\dagger(\omega^\prime)-\rho A_\mu^\dagger(\omega^\prime)A_\nu(\omega)\right).
\end{split}
\end{equation}
Here, we approximate the coefficients composed of $\gamma(\omega)$ and $\eta(\omega)$ as follows:
\begin{align}
  \frac{\gamma_{\mu\nu}(\omega)+\gamma_{\mu\nu}(\omega^\prime)}{2}\simeq& \sum_\lambda \gamma^{1/2}_{\mu\lambda}(\omega)\gamma^{1/2}_{\lambda\nu}(\omega^\prime),\\\label{eq:etaapprox}
  \eta_{\mu\nu}(\omega),\eta_{\mu\nu}(\omega^\prime)\simeq& \eta_{\mu\nu}\left(\frac{\omega+\omega^\prime}{2}\right).
\end{align}
The square root $\gamma_{\mu\nu}^{1/2}$ of the power spectrum is defined as 
\begin{equation}
  \gamma_{\mu\nu}(\omega)=\sum_\lambda \gamma^{1/2}_{\mu\lambda}(\omega)\gamma^{1/2}_{\lambda\nu}(\omega).
\end{equation}
Since we approximate the power spectrum using its square root, this approach to deriving the NRE is called the ``$\sqrt{\text{SD}}$-approach'' in Ref.~\cite{Davidovic2020completelypositive}, where the power spectrum is called the spectral density (SD).
This approximation is justified in the following two situations.
The first situation is that the condition $|\omega-\omega^\prime|\lesssim{\Delta t}^{-1}\ll{\tau_\mathrm{B}}^{-1}$ holds so that $\gamma_{\mu\nu}(\omega)\simeq\gamma_{\mu\nu}(\omega^\prime)$ and $\eta_{\mu\nu}(\omega)\simeq\eta_{\mu\nu}(\omega^\prime)$.
In fact,
\begin{equation}
  \begin{split}
    \gamma_{\mu\nu}(\omega)&=\int^{\infty}_{-\infty}dt e^{i\omega t}C_{\mu\nu}(t)\\
    &\simeq \int^{\sqrt{\tau_\mathrm{B}\Delta t}}_{-\sqrt{\tau_\mathrm{B}\Delta t}}dt e^{i\omega t}C_{\mu\nu}(t)~\text{($C(t)\sim0$ for $t>\sqrt{\tau_\mathrm{B}\Delta t}\gg\tau_\mathrm{B}$)}\\
    &= \int^{\sqrt{\tau_\mathrm{B}\Delta t}}_{-\sqrt{\tau_\mathrm{B}\Delta t}}dt e^{i\omega^\prime t}e^{i(\omega-\omega^\prime) t}C_{\mu\nu}(t)\\
    &\simeq\int^{\sqrt{\tau_\mathrm{B}\Delta t}}_{-\sqrt{\tau_\mathrm{B}\Delta t}}dt e^{i\omega^\prime t}C_{\mu\nu}(t)~\text{($e^{i(\omega-\omega^\prime)t}\simeq 1$ because $|(\omega-\omega^\prime)t|\ll1$ at $|t|\ll \Delta t$.)}\\
    &\simeq \int^{\infty}_{-\infty}dt e^{i\omega^\prime t}C_{\mu\nu}(t)=\gamma_{\mu\nu}(\omega^\prime).
  \end{split}
\end{equation}
Since $\gamma_{\mu\nu}(\omega)\simeq\gamma_{\mu\nu}(\omega^\prime)$ holds, $\eta_{\mu\nu}(\omega)\simeq\eta_{\mu\nu}(\omega^\prime)$ also holds for $|\omega-\omega^\prime|\lesssim{\Delta t}^{-1}\ll{\tau_\mathrm{B}}^{-1}$ and the approximation in Eq.~(\ref{eq:etaapprox}) is also justified.
The second situation is that $|\omega-\omega^\prime|\gtrsim {\Delta t}^{-1}$ holds so that the effect of such terms is canceled out after the coarse graining of time in the scale of $\Delta t$.
In this sense, the dynamics described by the obtained GKSL equation does not deviate from that described by the Redfield equation up to the coarse graining of time.

With this approximation, Eq.~(\ref{eq:beforeULE}) can be transformed as
\begin{equation}
  \label{eq:preULE}
  \begin{split}
    \frac{d}{dt}\rho=&-i[H_\mathrm{S},\rho ]+\sum_{\lambda}\sum_{\mu,\nu}\sum_{\omega,\omega^\prime}\gamma^{1/2}_{\mu\lambda}(\omega)\gamma^{1/2}_{\lambda\nu}(\omega^\prime)\left(A_\nu(\omega^\prime)\rho A_\mu^\dagger(\omega)-\frac{1}{2}\{A_\mu^\dagger(\omega)A_\nu(\omega^\prime),\rho \}\right)\\
    &+i\sum_{\mu,\nu}\sum_{\omega,\omega^\prime}\left(\eta_{\mu\nu}\left(\frac{\omega+\omega^\prime}{2}\right)+i\frac{\gamma_{\mu\nu}(\omega)-\gamma_{\mu\nu}(\omega^\prime)}{4}\right)\left(A_\mu^\dagger(\omega)A_\nu(\omega^\prime)\rho -\rho A_\mu^\dagger(\omega)A_\nu(\omega^\prime)\right).
\end{split}
\end{equation}
By defining the Lindblad operators $L_\lambda$ and the Lamb-shift Hamiltonian $H_{\mathrm{LS}}$ as
\begin{align}
  \label{eq:LindopApp}
  L_\lambda&=\sum_\nu\sum_\omega \gamma^{1/2}_{\lambda\nu}(\omega)A_\nu(\omega),\\
  H_\mathrm{LS}&=\sum_{\mu,\nu}\sum_{\omega,\omega^\prime}\left(\eta_{\mu\nu}\left(\frac{\omega+\omega^\prime}{2}\right)+i\frac{\gamma_{\mu\nu}(\omega)-\gamma_{\mu\nu}(\omega^\prime)}{4}\right)A_\mu^\dagger(\omega)A_\nu(\omega^\prime),
\end{align}
Eq.~(\ref{eq:preULE}) is written in the following GKSL form:
\begin{equation}
  \frac{d}{dt}\rho=-i[H_\mathrm{S}+H_\mathrm{LS},\rho ]+\sum_{\lambda=1}^k\left(L_\lambda\rho  L_\lambda^\dagger -\frac{1}{2}\{L_\lambda^\dagger L_\lambda,\rho \}\right).
\end{equation}

According to the derivation in Ref.~\cite{PhysRevB.102.115109}, the Lindblad operators of the NRE are given by
\begin{equation}
  \label{eq:ULELindop}
  L_\lambda=\sum_\nu\int^\infty_{-\infty}ds g_{\lambda\nu}(-s)A_\nu(s),
\end{equation}
where $g_{\mu\nu}(s)$ is defined as the Fourier transformation of $\gamma_{\mu\nu}^{1/2}(\omega)$:
\begin{equation}
  g_{\lambda\nu}(s)=\frac{1}{2\pi}\int^\infty_{-\infty}\gamma_{\lambda\nu}^{1/2}(\omega)e^{-i\omega s}d\omega.
\end{equation}
The Lindblad operators in Eq.~(\ref{eq:ULELindop}) are equivalent to those in Eq.~(\ref{eq:LindopApp}) since
\begin{equation}
  \begin{split}
    L_\lambda&=\int^\infty_{-\infty}ds \sum_{\nu} g_{\lambda\nu}(-s) A_\nu(s)\\\label{eq:78}
    &=\sum_{m,n}\int^\infty_{-\infty}ds \sum_{\nu} g_{\lambda\nu}(-s) e^{i(E_n-E_m)s}\bra{E_m}A_\nu\ket{E_n}\ket{E_m}\bra{E_n}\\
    &=\sum_{\omega}\int^\infty_{-\infty}ds \sum_{\nu} g_{\lambda\nu}(-s) e^{i\omega s}A_\nu(\omega)\\
    &=\sum_{\omega}\sum_\nu\gamma^{1/2}_{\lambda\nu}(\omega)A_\nu(\omega).
  \end{split}
\end{equation}


By writing the Lindblad operators of the NRE as in Eq.~(\ref{eq:ULELindop}),
we can also derive the local NRE by replacing $A_\mu(s)$ with $A_\mu^\mathrm{loc}(s)$ in the Lindblad operators~(\ref{eq:ULELindop}).
In this sense the local NRE can also be considered as the approximation of the original NRE.
Therefore, the local NRE becomes more accurate for a larger subsystem $\Omega_\mu$ in contrast to the local Davies equation.
However, the computational costs of calculating the Lindblad operators and the Lamb-shift Hamiltonian increase as the size of the subsystem $\Omega_\mu$ increases.

\section{Efficient calculation of the largest eigenvalue of a dissipator}
\label{sec:appendixBB}
In this appendix, we show that the largest eigenvalue of a dissipator of the quadratic open fermionic system can efficiently be calculated.
In the quadratic open fermionic system, the QME can be represented as in Eq.~(\ref{eq:quadraticGKSL}).
The problem of finding the steady state of this system is then reduced to diagonalizing the matrix $A$~\cite{Prosen_2008,Prosen_2010PT}, whose nonzero components are given by
\begin{equation}
  \label{eq:Amatrix}
  \begin{aligned}
    A_{2m-1,2n-1}&=-2iH_{m,n}-M_{m,n}+M_{n,m},&A_{2m,2n}&=-2iH_{m,n}+M_{m,n}-M_{n,m},\\
    A_{2m-1,2n}&=iM_{n,m}+iM_{m,n},&A_{2m,2n-1}&=-iM_{m,n}-iM_{n,m}.
  \end{aligned}
\end{equation}
Here, $H=(H_{m,n})$ is an antisymmetric Hermitian matrix which satisfies $H_\mathrm{S}=\sum_{m,n}H_{m,n}w_mw_n$.
Since $A$ is antisymmetric, eigenvalues of $A$ are expressed as $\beta_j,-\beta_j,~(j=1,\ldots,2L)$, where $\Re\beta_j\geq0$.

Let $\mathcal{L}[\rho]$ be given by the right-hand side of Eq.~(\ref{eq:quadraticGKSL}).
We assume that the real part of each eigenvalue of $\mathcal{L}$ is negative except for a single zero eigenvalue, which corresponds to a steady state.
Here, we focus on the dynamics in subspace $\mathcal{K}^+$ composed of an even number of fermionic operators $w_m$ since we are interested in the expectation values of the product of the even number of $w_m$'s such as $W_{mn}=\langle w_mw_n\rangle$.
For the precise definition of $\mathcal{K}^+$, see Refs.~\cite{Prosen_2008,Prosen_2010PT}.
Let $\mathcal{P}_+$ be a projector onto $\mathcal{K}^+$.
The dynamics in subspace $\mathcal{K}^+$ is described by $\mathcal{L}_+\coloneqq\mathcal{P}_+\mathcal{L}\mathcal{P}_+$.
Note that the generator $\mathcal{L}$ is block diagonalized as $\mathcal{L}=\mathcal{P}_+\mathcal{L}\mathcal{P}_++(1-\mathcal{P}_+)\mathcal{L}(1-\mathcal{P}_+)$~\cite{Prosen_2008,Prosen_2010PT}.

By assuming that the real part of the nonzero eigenvalues of $\mathcal{L}$ is negative and that $\mathcal{L}$ has a unique zero eigenvalue, $\mathcal{L}$ is expressed as
\begin{equation}
  \mathcal{L}_+=-2\sum_i\beta_ib_i^\prime b_i,
\end{equation}
where $b_i^\prime$ and $b_i$ play a similar role as the creation and annihilation operators of normal modes and satisfy the anticommutation relations $\{b^\prime_i,b_j\}=\delta_{ij},\{b_i,b_j\}=\{b^\prime_i,b^\prime_j\}=0$.
The steady state $\rho_\mathrm{steady}$ of $\mathcal{L}$ is given by the ``vacuum state'' satisfying $b_i\rho_\mathrm{steady}=0$ for all $i$.
See Ref.~\cite{Prosen_2008} for the details.
Therefore, the eigenvalues $\lambda_\nu$ of $\mathcal{L}_+$ are given by all the possible binary linear combinations of $\beta_j$'s:
\begin{equation}
  \lambda_\nu=-2\sum_jv_{\nu,j}\beta_j,~v_{\nu,j}\in\{0,1\}.
\end{equation}
To evaluate $\tau_{\mathrm{SB}}$, we need to have the eigenvalues of the dissipator $\mathcal{D}$ defined in Eq.~(\ref{eq:dissipator}). They are obtained by setting $H_{m,n}=0$ in Eq.~(\ref{eq:Amatrix}).
For the GKSL equation, $M_{m,n}$ is a Hermitian matrix. Therefore, when $H_{m,n}=0$, the matrix $A$ corresponding to $\mathcal{D}$ is also a Hermitian matrix and all the eigenvalues of $A$ are real.
Since $\Re\beta_j\geq 0$, the eigenvalue $\lambda_\mathrm{max}$ with the largest absolute value of $\mathcal{D}_+\coloneqq\mathcal{P}_+\mathcal{D}\mathcal{P}_+$ is given by
\begin{equation}
  \lambda_\mathrm{max}=-2\sum_j\beta_j,
\end{equation}
if the real part of each nonzero eigenvalue of $\mathcal{D}$ is negative.
Thus, the eigenvalue with the largest absolute value of a dissipator of QMEs can be calculated efficiently by using the sum rule shown in Ref.~\cite{Prosen_2008}:
\begin{equation}
  2\sum_j\beta_j=2\tr M.
\end{equation}

\section{Additional numerical results}
\label{sec:appendixB}
\begin{figure}
  \centering
  \includegraphics[width=12 cm]{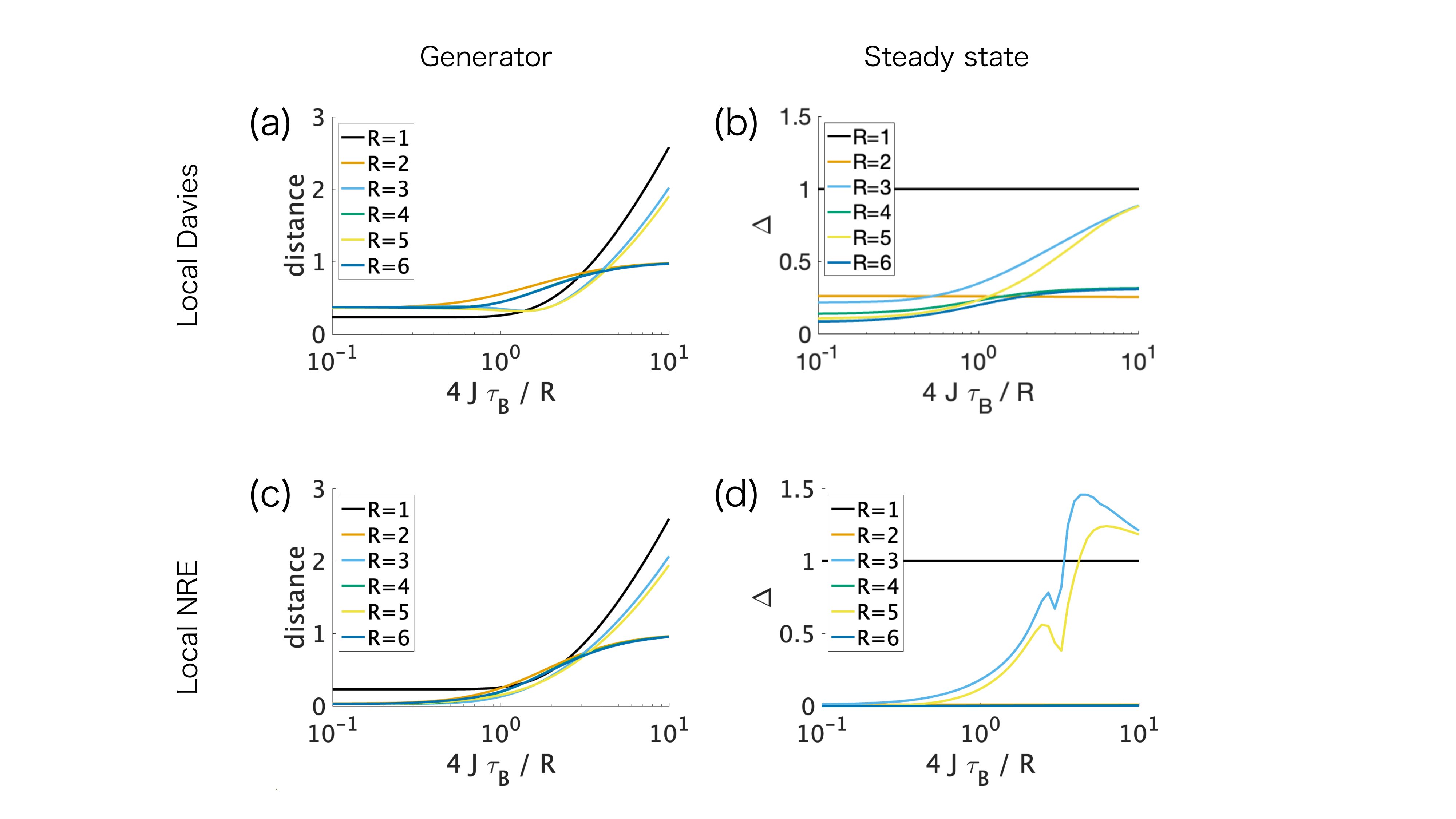}
  \caption{(a, c) Distance between generators of the local GKSL equations and that of the Redfield equation. (b,d) deviations in the steady states of the local Davies equation and the local NRE.
  The parameters are set to be $L=128$, $N=16$, $\tau_\mathrm{SB}=100$, $J=1$ and $\omega_0=0$. The inverse temperatures of the baths are set to be $\beta_r=0.5$ and $\beta_l=0.1$.
  When the relaxation time $\tau_\mathrm{B}$ is small, both the distance and the deviation become small regardless of whether $R$ is even or odd except for the case of $R=1$.
  When $\tau_\mathrm{B}$ is large, the deviation tends to be large for odd $R$.
  (a) Distance between the generator of the local Davies equation and that of the Redfield equation.
  (b) Deviation of the steady state of the local Davies equation from that of the Redfield equation.
  (c) Distance between the generator of the local NRE and that of the Redfield equation.
  (d) Deviation of the steady state of the local NRE from that of the Redfield equation. The deviations for even $R$ are negligible compared with those for odd $R$. The deviation is larger around $\tau_\mathrm{B}=1$ when $R$ is odd while the distances of the generators are comparable regardless of whether $R$ is even or odd as seen in (c).
  }
  \label{fig:odd}
\end{figure}

Here we discuss the deviations in the generators and the steady state when the radius $R$ of the subsystem is given by an odd integer.
Compared with the case of even $R$, the distance between generators for odd $R$ becomes larger in the region where $\tau_\mathrm{B}$ is large for both the local Davies equation and the local NRE (see Figs.~\ref{fig:odd} (a) and (c)).
This makes the deviation in the steady state for odd $R$ larger in the region where $\tau_\mathrm{B}$ is large (see Figs.~\ref{fig:odd} (b) and (d)).
If $\tau_\mathrm{B}$ is sufficiently small, the behavior of the deviations does not significantly depend on whether $R$ is even or odd.
Since the local GKSL equation is valid when $\tau_\mathrm{B}$ is smaller than $R/\zeta_0$ (where $\zeta_0=4J$), the even-odd dependence on $R$ in the deviation for large $\tau_\mathrm{B}$ is not relevant to the validity of the local GKSL equation.

The deviation of the steady state of the local GKSL equation with $R=1$ from that of the Redfield equation is always equal to one because the steady state of the local GKSL equation with $R=1$ is a product state and the off-diagonal components of $W$ vanish when $\omega_0=0$.
Since the diagonal components of $W$ are always one, the deviation $\Delta$ is given by
\begin{equation}
  \Delta\coloneqq \frac{\max_{kl}|W_{kl}^{\mathrm{locGKSL}}-W_{kl}^{\mathrm{Redfield}}|}{\max_{k\neq l}W_{kl}^{\mathrm{Redfield}}} =\frac{\max_{k\neq l}|W_{kl}^{\mathrm{Redfield}}|}{\max_{k\neq l}W_{kl}^{\mathrm{Redfield}}}=1.
\end{equation}

\twocolumngrid

\bibliography{aaa}

\begin{thebibliography}{10}

\bibitem{breuer2002theory}
Heinz-Peter Breuer, Francesco Petruccione, et~al.
\newblock {\em The theory of open quantum systems}.
\newblock Oxford University Press on Demand, 2002.

\bibitem{AJDaley}
Andrew~J. Daley.
\newblock Quantum trajectories and open many-body quantum systems.
\newblock {\em Advances in Physics}, 63(2):77--149, 2014.

\bibitem{MULLER20121}
Markus Müller, Sebastian Diehl, Guido Pupillo, and Peter Zoller.
\newblock Engineered open systems and quantum simulations with atoms and ions.
\newblock In Paul Berman, Ennio Arimondo, and Chun Lin, editors, {\em Advances in Atomic, Molecular, and Optical Physics}, volume~61 of {\em Advances In Atomic, Molecular, and Optical Physics}, pages 1--80. Academic Press, 2012.

\bibitem{10.1143/PTP.20.948}
Sadao Nakajima.
\newblock {On Quantum Theory of Transport Phenomena: Steady Diffusion}.
\newblock {\em Progress of Theoretical Physics}, 20(6):948--959, 12 1958.

\bibitem{doi:10.1063/1.1731409}
Robert Zwanzig.
\newblock Ensemble method in the theory of irreversibility.
\newblock {\em The Journal of Chemical Physics}, 33(5):1338--1341, 1960.

\bibitem{PhysRev.89.728}
R.~K. Wangsness and F.~Bloch.
\newblock The dynamical theory of nuclear induction.
\newblock {\em Phys. Rev.}, 89:728--739, Feb 1953.

\bibitem{5392713}
A.~G. Redfield.
\newblock On the theory of relaxation processes.
\newblock {\em IBM Journal of Research and Development}, 1(1):19--31, 1957.

\bibitem{REDFIELD19651}
A.G. Redfield.
\newblock The theory of relaxation processes.
\newblock In John~S. Waugh, editor, {\em Advances in Magnetic Resonance}, volume~1 of {\em Advances in Magnetic and Optical Resonance}, pages 1--32. Academic Press, 1965.

\bibitem{doi:10.1063/1.522979}
Vittorio Gorini, Andrzej Kossakowski, and E.~C.~G. Sudarshan.
\newblock Completely positive dynamical semigroups of n‐level systems.
\newblock {\em Journal of Mathematical Physics}, 17(5):821--825, 1976.

\bibitem{Lindblad1976}
G.~Lindblad.
\newblock On the generators of quantum dynamical semigroups.
\newblock {\em Communications in Mathematical Physics}, 48(2):119--130, 1976.

\bibitem{barreiro2011open}
Julio~T Barreiro, Markus M{\"u}ller, Philipp Schindler, Daniel Nigg, Thomas Monz, Michael Chwalla, Markus Hennrich, Christian~F Roos, Peter Zoller, and Rainer Blatt.
\newblock An open-system quantum simulator with trapped ions.
\newblock {\em Nature}, 470(7335):486--491, 2011.

\bibitem{PhysRevLett.110.035302}
G.~Barontini, R.~Labouvie, F.~Stubenrauch, A.~Vogler, V.~Guarrera, and H.~Ott.
\newblock Controlling the dynamics of an open many-body quantum system with localized dissipation.
\newblock {\em Phys. Rev. Lett.}, 110:035302, Jan 2013.

\bibitem{PhysRevX.7.011034}
Henrik~P. L\"uschen, Pranjal Bordia, Sean~S. Hodgman, Michael Schreiber, Saubhik Sarkar, Andrew~J. Daley, Mark~H. Fischer, Ehud Altman, Immanuel Bloch, and Ulrich Schneider.
\newblock Signatures of many-body localization in a controlled open quantum system.
\newblock {\em Phys. Rev. X}, 7:011034, Mar 2017.

\bibitem{doi:10.1126/sciadv.1701513}
Takafumi Tomita, Shuta Nakajima, Ippei Danshita, Yosuke Takasu, and Yoshiro Takahashi.
\newblock Observation of the {M}ott insulator to superfluid crossover of a driven-dissipative {B}ose-{H}ubbard system.
\newblock {\em Science Advances}, 3(12):e1701513, 2017.

\bibitem{bouganne2020anomalous}
Rapha{\"e}l Bouganne, Manel Bosch~Aguilera, Alexis Ghermaoui, J{\'e}r{\^o}me Beugnon, and Fabrice Gerbier.
\newblock Anomalous decay of coherence in a dissipative many-body system.
\newblock {\em Nature Physics}, 16(1):21--25, 2020.

\bibitem{PhysRevX.11.041046}
Francesco Ferri, Rodrigo Rosa-Medina, Fabian Finger, Nishant Dogra, Matteo Soriente, Oded Zilberberg, Tobias Donner, and Tilman Esslinger.
\newblock Emerging dissipative phases in a superradiant quantum gas with tunable decay.
\newblock {\em Phys. Rev. X}, 11:041046, Dec 2021.

\bibitem{PhysRevB.80.035110}
Giuliano Benenti, Giulio Casati, Toma\ifmmode \check{z}\else~\v{z}\fi{} Prosen, Davide Rossini, and Marko \ifmmode \check{Z}\else \v{Z}\fi{}nidari\ifmmode~\check{c}\else \v{c}\fi{}.
\newblock Charge and spin transport in strongly correlated one-dimensional quantum systems driven far from equilibrium.
\newblock {\em Phys. Rev. B}, 80:035110, Jul 2009.

\bibitem{PhysRevLett.106.217206}
Toma{\ifmmode \check{z}\else \v{z}\fi{}} Prosen.
\newblock Open {XXZ} spin chain: Nonequilibrium steady state and a strict bound on ballistic transport.
\newblock {\em Phys. Rev. Lett.}, 106:217206, May 2011.

\bibitem{RevModPhys.94.045006}
Gabriel~T. Landi, Dario Poletti, and Gernot Schaller.
\newblock Nonequilibrium boundary-driven quantum systems: Models, methods, and properties.
\newblock {\em Rev. Mod. Phys.}, 94:045006, Dec 2022.

\bibitem{michel2003fourier}
Mathias Michel, Michael Hartmann, Jochen Gemmer, and Guenter Mahler.
\newblock Fourier's law confirmed for a class of small quantum systems.
\newblock {\em The European Physical Journal B-Condensed Matter and Complex Systems}, 34(3):325--330, 2003.

\bibitem{mejia2007heat}
Carlos Mejia-Monasterio and Hannu Wichterich.
\newblock Heat transport in quantum spin chains.
\newblock {\em The European Physical Journal Special Topics}, 151(1):113--125, 2007.

\bibitem{Steinigeweg_2009}
R.~Steinigeweg, M.~Ogiewa, and J.~Gemmer.
\newblock Equivalence of transport coefficients in bath-induced and dynamical scenarios.
\newblock {\em Europhysics Letters}, 87(1):10002, jul 2009.

\bibitem{PhysRevE.86.061118}
Daniel Manzano, Markus Tiersch, Ali Asadian, and Hans~J. Briegel.
\newblock Quantum transport efficiency and {F}ourier's law.
\newblock {\em Phys. Rev. E}, 86:061118, Dec 2012.

\bibitem{Manzano_2016}
Daniel Manzano, Chern Chuang, and Jianshu Cao.
\newblock Quantum transport in d-dimensional lattices.
\newblock {\em New J. Phys.}, 18(4):043044, apr 2016.

\bibitem{PhysRevE.76.031115}
Hannu Wichterich, Markus~J. Henrich, Heinz-Peter Breuer, Jochen Gemmer, and Mathias Michel.
\newblock Modeling heat transport through completely positive maps.
\newblock {\em Phys. Rev. E}, 76:031115, Sep 2007.

\bibitem{PhysRevB.105.115139}
G.~Schaller, F.~Queisser, N.~Szpak, J.~K\"onig, and R.~Sch\"utzhold.
\newblock Environment-induced decay dynamics of antiferromagnetic order in mott-hubbard systems.
\newblock {\em Phys. Rev. B}, 105:115139, Mar 2022.

\bibitem{PhysRevLett.107.137201}
Toma\ifmmode \check{z}\else~\v{z}\fi{} Prosen.
\newblock Exact nonequilibrium steady state of a strongly driven open $xxz$ chain.
\newblock {\em Phys. Rev. Lett.}, 107:137201, Sep 2011.

\bibitem{RevModPhys.93.025003}
B.~Bertini, F.~Heidrich-Meisner, C.~Karrasch, T.~Prosen, R.~Steinigeweg, and M.~\ifmmode \check{Z}\else \v{Z}\fi{}nidari\ifmmode~\check{c}\else \v{c}\fi{}.
\newblock Finite-temperature transport in one-dimensional quantum lattice models.
\newblock {\em Rev. Mod. Phys.}, 93:025003, May 2021.

\bibitem{PhysRevLett.104.190401}
David Poulin.
\newblock Lieb-robinson bound and locality for general markovian quantum dynamics.
\newblock {\em Phys. Rev. Lett.}, 104:190401, May 2010.

\bibitem{PhysRevE.101.042116}
Tatsuhiko Shirai and Takashi Mori.
\newblock Thermalization in open many-body systems based on eigenstate thermalization hypothesis.
\newblock {\em Phys. Rev. E}, 101:042116, Apr 2020.

\bibitem{PhysRevA.105.032208}
Devashish Tupkary, Abhishek Dhar, Manas Kulkarni, and Archak Purkayastha.
\newblock Fundamental limitations in lindblad descriptions of systems weakly coupled to baths.
\newblock {\em Phys. Rev. A}, 105:032208, Mar 2022.

\bibitem{PhysRevA.107.062216}
Devashish Tupkary, Abhishek Dhar, Manas Kulkarni, and Archak Purkayastha.
\newblock Searching for lindbladians obeying local conservation laws and showing thermalization.
\newblock {\em Phys. Rev. A}, 107:062216, Jun 2023.

\bibitem{Levy_2014}
Amikam Levy and Ronnie Kosloff.
\newblock The local approach to quantum transport may violate the second law of thermodynamics.
\newblock {\em Europhysics Letters}, 107(2):20004, jul 2014.

\bibitem{Barra2015}
Felipe Barra.
\newblock The thermodynamic cost of driving quantum systems by their boundaries.
\newblock {\em Scientific Reports}, 5:14873, Oct 2015.

\bibitem{Hofer_2017}
Patrick~P Hofer, Martí Perarnau-Llobet, L~David~M Miranda, Géraldine Haack, Ralph Silva, Jonatan~Bohr Brask, and Nicolas Brunner.
\newblock Markovian master equations for quantum thermal machines: local versus global approach.
\newblock {\em New J. Phys.}, 19(12):123037, dec 2017.

\bibitem{PhysRevE.97.022115}
Emmanuel Pereira.
\newblock Heat, work, and energy currents in the boundary-driven $xxz$ spin chain.
\newblock {\em Phys. Rev. E}, 97:022115, Feb 2018.

\bibitem{DeChiara_2018}
Gabriele~De Chiara, Gabriel Landi, Adam Hewgill, Brendan Reid, Alessandro Ferraro, Augusto~J Roncaglia, and Mauro Antezza.
\newblock Reconciliation of quantum local master equations with thermodynamics.
\newblock {\em New Journal of Physics}, 20(11):113024, nov 2018.

\bibitem{RevModPhys.93.035008}
Gabriel~T. Landi and Mauro Paternostro.
\newblock Irreversible entropy production: From classical to quantum.
\newblock {\em Rev. Mod. Phys.}, 93:035008, Sep 2021.

\bibitem{Scali2021localmaster}
Stefano Scali, Janet Anders, and Luis~A. Correa.
\newblock Local master equations bypass the secular approximation.
\newblock {\em {Quantum}}, 5:451, May 2021.

\bibitem{PhysRevResearch.4.013171}
Michael Konopik and Eric Lutz.
\newblock Local master equations may fail to describe dissipative critical behavior.
\newblock {\em Phys. Rev. Res.}, 4:013171, Mar 2022.

\bibitem{PhysRev.129.1880}
Jayaseetha Rau.
\newblock Relaxation phenomena in spin and harmonic oscillator systems.
\newblock {\em Phys. Rev.}, 129:1880--1888, Feb 1963.

\bibitem{Englert2002}
Berthold-Georg Englert and Giovanna Morigi.
\newblock {\em Five Lectures on Dissipative Master Equations}, pages 55--106.
\newblock Springer Berlin Heidelberg, Berlin, Heidelberg, 2002.

\bibitem{PhysRevA.65.040102}
Giovanna Morigi, Enrique Solano, Berthold-Georg Englert, and Herbert Walther.
\newblock Measuring irreversible dynamics of a quantum harmonic oscillator.
\newblock {\em Phys. Rev. A}, 65:040102, Apr 2002.

\bibitem{PhysRevLett.88.097905}
Valerio Scarani, M\'ario Ziman, Peter \ifmmode \check{S}\else \v{S}\fi{}telmachovi\ifmmode~\check{c}\else \v{c}\fi{}, Nicolas Gisin, and Vladim\'{\i}r Bu\ifmmode~\check{z}\else \v{z}\fi{}ek.
\newblock Thermalizing quantum machines: Dissipation and entanglement.
\newblock {\em Phys. Rev. Lett.}, 88:097905, Feb 2002.

\bibitem{PhysRevLett.102.207207}
Dragi Karevski and Thierry Platini.
\newblock Quantum nonequilibrium steady states induced by repeated interactions.
\newblock {\em Phys. Rev. Lett.}, 102:207207, May 2009.

\bibitem{Ciccarello+2017+53+63}
Francesco Ciccarello.
\newblock Collision models in quantum optics.
\newblock {\em Quantum Measurements and Quantum Metrology}, 4(1):53--63, 2017.

\bibitem{CICCARELLO20221}
Francesco Ciccarello, Salvatore Lorenzo, Vittorio Giovannetti, and G.~Massimo Palma.
\newblock Quantum collision models: Open system dynamics from repeated interactions.
\newblock {\em Physics Reports}, 954:1--70, 2022.
\newblock Quantum collision models: Open system dynamics from repeated interactions.

\bibitem{Trushechkin_2016}
A.~S. Trushechkin and I.~V. Volovich.
\newblock Perturbative treatment of inter-site couplings in the local description of open quantum networks.
\newblock {\em Europhysics Letters}, 113(3):30005, feb 2016.

\bibitem{schnell2023global}
Alexander Schnell.
\newblock Global becomes local: Efficient many-body dynamics for global master equations, 2023.

\bibitem{Lieb1972}
E.H. Lieb and D.W. Robinson.
\newblock The finite group velocity of quantum spin systems.
\newblock {\em Communications in Mathematical Physics}, 28:251--257, Sep 1972.

\bibitem{doi:10.1137/18M1231511}
Jeongwan Haah, Matthew~B. Hastings, Robin Kothari, and Guang~Hao Low.
\newblock Quantum algorithm for simulating real time evolution of lattice {H}amiltonians.
\newblock {\em SIAM Journal on Computing}, 0(0):FOCS18--250--FOCS18--284, 0.

\bibitem{Davies1974}
E.~B. Davies.
\newblock Markovian master equations.
\newblock {\em Communications in Mathematical Physics}, 39(2):91--110, 1974.

\bibitem{doi:10.1063/1.4907370}
Jan Jeske, David~J. Ing, Martin~B. Plenio, Susana~F. Huelga, and Jared~H. Cole.
\newblock {B}loch-{R}edfield equations for modeling light-harvesting complexes.
\newblock {\em The Journal of Chemical Physics}, 142(6):064104, 2015.

\bibitem{https://doi.org/10.48550/arxiv.1710.09939}
J~D Cresser and C~Facer.
\newblock Coarse-graining in the derivation of {M}arkovian master equations and its significance in quantum thermodynamics, 2017.

\bibitem{PhysRevA.100.012107}
Donato Farina and Vittorio Giovannetti.
\newblock Open-quantum-system dynamics: Recovering positivity of the {R}edfield equation via the partial secular approximation.
\newblock {\em Phys. Rev. A}, 100:012107, Jul 2019.

\bibitem{Cohen-Tannoudji_1986}
Claude Cohen-Tannoudji.
\newblock Fluctuations in radiative processes.
\newblock {\em Physica Scripta}, 1986(T12):19, jan 1986.

\bibitem{LIDAR200135}
Daniel~A. Lidar, Zsolt Bihary, and K.Birgitta Whaley.
\newblock From completely positive maps to the quantum {M}arkovian semigroup master equation.
\newblock {\em Chemical Physics}, 268(1):35--53, 2001.

\bibitem{PhysRevA.78.022106}
Gernot Schaller and Tobias Brandes.
\newblock Preservation of positivity by dynamical coarse graining.
\newblock {\em Phys. Rev. A}, 78:022106, Aug 2008.

\bibitem{PhysRevA.79.032110}
Gernot Schaller, Philipp Zedler, and Tobias Brandes.
\newblock Systematic perturbation theory for dynamical coarse-graining.
\newblock {\em Phys. Rev. A}, 79:032110, Mar 2009.

\bibitem{PhysRevA.88.012103}
Christian Majenz, Tameem Albash, Heinz-Peter Breuer, and Daniel~A. Lidar.
\newblock Coarse graining can beat the rotating-wave approximation in quantum {M}arkovian master equations.
\newblock {\em Phys. Rev. A}, 88:012103, Jul 2013.

\bibitem{Mozgunov2020completelypositive}
Evgeny Mozgunov and Daniel Lidar.
\newblock Completely positive master equation for arbitrary driving and small level spacing.
\newblock {\em {Quantum}}, 4:227, February 2020.

\bibitem{PhysRevB.97.035432}
Gediminas Kir{\ifmmode \check{s}\else \v{s}\fi{}}anskas, Martin Francki\'e, and Andreas Wacker.
\newblock Phenomenological position and energy resolving {L}indblad approach to quantum kinetics.
\newblock {\em Phys. Rev. B}, 97:035432, Jan 2018.

\bibitem{PhysRevB.102.115109}
Frederik Nathan and Mark~S. Rudner.
\newblock Universal {L}indblad equation for open quantum systems.
\newblock {\em Phys. Rev. B}, 102:115109, Sep 2020.

\bibitem{Davidovic2020completelypositive}
Dragomir Davidovi{\'{c}}.
\newblock Completely {P}ositive, {S}imple, and {P}ossibly {H}ighly {A}ccurate {A}pproximation of the {R}edfield {E}quation.
\newblock {\em {Quantum}}, 4:326, September 2020.

\bibitem{PhysRevE.104.014110}
Tobias Becker, Ling-Na Wu, and Andr\'e Eckardt.
\newblock Lindbladian approximation beyond ultraweak coupling.
\newblock {\em Phys. Rev. E}, 104:014110, Jul 2021.

\bibitem{PhysRevA.102.032207}
V.~Yu. Shishkov, E.~S. Andrianov, A.~A. Pukhov, A.~P. Vinogradov, and A.~A. Lisyansky.
\newblock Perturbation theory for {L}indblad superoperators for interacting open quantum systems.
\newblock {\em Phys. Rev. A}, 102:032207, Sep 2020.

\bibitem{doi:10.1146/annurev-conmatphys-040721-015537}
Takashi Mori.
\newblock Floquet states in open quantum systems.
\newblock {\em Annual Review of Condensed Matter Physics}, 14(1):35--56, 2023.

\bibitem{https://doi.org/10.48550/arxiv.2206.02917}
Frederik Nathan and Mark~S. Rudner.
\newblock High accuracy steady states obtained from the {U}niversal {L}indblad {E}quation, 2022.

\bibitem{PhysRevA.101.012103}
Richard Hartmann and Walter~T. Strunz.
\newblock Accuracy assessment of perturbative master equations: Embracing nonpositivity.
\newblock {\em Phys. Rev. A}, 101:012103, Jan 2020.

\bibitem{Prosen_2008}
Toma{\ifmmode \check{z}\else \v{z}\fi{}} Prosen.
\newblock Third quantization: a general method to solve master equations for quadratic open fermi systems.
\newblock {\em New J. Phys.}, 10(4):043026, apr 2008.

\bibitem{Prosen_2010}
Toma{\ifmmode \check{z}\else \v{z}\fi{}} Prosen and Thomas~H Seligman.
\newblock Quantization over boson operator spaces.
\newblock {\em Journal of Physics A: Mathematical and Theoretical}, 43(39):392004, sep 2010.

\bibitem{Prosen_2010PT}
Tomaž Prosen and Bojan Žunkovič.
\newblock Exact solution of markovian master equations for quadratic fermi systems: thermal baths, open xy spin chains and non-equilibrium phase transition.
\newblock {\em New J. Phys.}, 12(2):025016, feb 2010.

\bibitem{PhysRevA.95.052107}
Chu Guo and Dario Poletti.
\newblock Solutions for bosonic and fermionic dissipative quadratic open systems.
\newblock {\em Phys. Rev. A}, 95:052107, May 2017.

\bibitem{PhysRevB.109.L180408}
Federico Garcia-Gaitan and Branislav~K. Nikoli\ifmmode~\acute{c}\else \'{c}\fi{}.
\newblock Fate of entanglement in magnetism under lindbladian or non-markovian dynamics and conditions for their transition to landau-lifshitz-gilbert classical dynamics.
\newblock {\em Phys. Rev. B}, 109:L180408, May 2024.

\end{thebibliography}
\bibliographystyle{abbrv}
\bibliographystyle{unsrt}
\end{document}